\documentclass[structabstract]{aa}  

\def\hi{\relax \ifmmode {\mbox H\,{\scshape i}}\else H\,{\scshape i}\fi}
\def\hii{\relax \ifmmode {\mbox H\,{\scshape ii}}\else H\,{\scshape ii}\fi}
\def\nii{\relax \ifmmode {\mbox N\,{\scshape ii}}\else N\,{\scshape ii}\fi}
\def\oi{\relax \ifmmode {\mbox O\,{\scshape i}}\else O\,{\scshape i}\fi}
\def\oii{\relax \ifmmode {\mbox O\,{\scshape ii}}\else O\,{\scshape ii}\fi}
\def\oiii{\relax \ifmmode {\mbox O\,{\scshape iii}}\else O\,{\scshape iii}\fi}
\def\sii{\relax \ifmmode {\mbox S\,{\scshape ii}}\else S\,{\scshape ii}\fi}
\def\siii{\relax \ifmmode {\mbox S\,{\scshape iii}}\else S\,{\scshape iii}\fi}
\def\neiii{\relax \ifmmode {\mbox Ne\,{\scshape iii}}\else Ne\,{\scshape iii}\fi}
\def\cliii{\relax \ifmmode {\mbox Cl\,{\scshape iii}}\else Cl\,{\scshape iii}\fi}
\def\ha{\relax \ifmmode {\mbox H}\alpha\else H$\alpha$\fi}
\def\hb{\relax \ifmmode {\mbox H}\beta\else H$\beta$\fi}

\usepackage{graphicx}
\usepackage{lscape}
\usepackage{epsfig}
\usepackage{txfonts}
\begin{document}
   \title{Ionised gas abundances in barred spiral galaxies\footnote{Based on observations obtained at Siding Spring Observatory (RSAA, ANU, australia) and the INT telescope at the ING, La Palma, spain}}

   \subtitle{}

    \author{E. Florido\inst{1,}\inst{2}, 
          I. P\'erez 
          \inst{1,}\inst{2},
          A. Zurita
          \inst{1,}\inst{2}
             \and
          P. S\'anchez-Bl\'azquez \inst{3}
          }

   \institute{Dpto. de F\' isica Te\'orica y del Cosmos, University of Granada,
              Facultad de Ciencias (Edificio Mecenas), E-18071, Granada, Spain\\
              \email{estrella@ugr.es, isa@ugr.es}
         \and
             Instituto Universitario Carlos I de F\'isica Te\'orica y Computacional, Facultad de Ciencias, E-18071, Granada, Spain\\
             \email{azurita@ugr.es}
         \and
             Dpto. F\'isica Te\'orica, Universidad Aut\'onoma de Madrid, Cantoblanco, E-28049, Madrid, Spain\\
             \email{p.sanchezblazquez@uam.es}
             }

   \date{}

  \abstract
   {}
{This is the third paper of a series devoted to study the 
properties of bars from long slit spectroscopy  to understand 
their formation, evolution and their influence on the evolution 
of disk galaxies.
In this work we aim to determine the gas metallicity
distribution of a sample of 20 barred early-type galaxies.  We
compare the nebular and stellar metallicity distributions
to conclude about the origin of the warm gas.}
   {Long-slit spectroscopy along the bar was obtained 
   and metallicities derived using different calibrations. We compare the results of nebular emission
metallicities using different semi-empirical methods. We carry out AGN
diagnostic diagrams along the radius to determine the radius of
influence of the AGN and the nuclei nature of the studied galaxies. We
then derive the gas metallicities along the bars and compare the
results to the distribution of stellar metallicities  in the same
regions.}
{Most of the gas emission is centrally concentrated, although
15 galaxies also show emission along the bar. In the central regions,  gas
oxygen abundances are in the  range  12+$\log$(O/H)= 8.4-9.1. 
The nebular metallicity  gradients are
very shallow in the bulge and bar regions. For three galaxies (one of them a LINER), the
gas metallicities lie well below the stellar ones in the bulge region. These
results do not depend on the choice of the semi-empirical calibration used to calculate
the abundances. We see that the galaxies with the lowest abundances are those with the largest rotational velocities.
Unlike with the stellar
abundances, we find no correlation between the nebular abundances and the
central velocity dispersion. In most galaxies the slope 
for the nebular abundance distribution in the bulge region is shallower than that for the stellar metallicity.}
   {The presence of gas of significantly lower metallicity than the
stellar abundances in three of our galaxies, points to an external origin as the source of the
gas that fuels the present star formation in the centre of some
early-type barred galaxies. The fact that the bar/disk nebular
metallicities are higher than the central ones might be indicating that
the gas could be accreted via cooling flows instead of radial accretion
from gas sitting in the outer parts of the disk.}
   \keywords{Galaxies: abundances --Galaxies: evolution
                -- Galaxies: spiral}
   \maketitle
%
\section{Introduction}
 
With the exception of few primordial light elements, stellar nucleosynthesis is responsible of 
the secular metal enrichment in galaxies. 
The final metallicity distribution within a galaxy is reasonably well reproduced
by chemical evolution models considering the appropiate star formation history  together
with episodes of gaseous inflow and outflows (e.g. Portinari \& Chiosi 1999; Chiappini et al. 2000; 
Prantzos 2008, Colavitti et al. 2009). However, 
the metallicity gradients can be futher
modified by dynamical processes (e.g., Roskar et al. 2008, Sch\"onric \& Binney 2009; S\'anchez-Bl\'azquez et al.
2009). It is, therefore, evident that observing how metals are distributed 
in a galaxy should highly constrain its evolution. However, the detailed dynamical  processes  responsible 
for the modification of the metallicity distribution and their importance  are not yet well understood.

In particular, it is not known, from the observational point of view, the importance that bars
might have in producing the metallicity radial mixing. Bars are believed to affect the overall dynamics 
of the galaxy and are a well known mechanism to induce secular evolution (Athanassoula 2003; 
Pfenniger \& Friedli 1991). For instance,  in a previous work (P\'erez \& S\'anchez-Bl\'azquez 2011, 
hereinafter, Paper II) we hinted that the bulges of early-type barred galaxies show different stellar enrichment 
histories compared to the bulges of their unbarred counterparts (see also Ellison et al. 2001 for a similar conclusion for gas-phase metallicities).

Gas and stars respond to the gravitational potential due to their own nature, being viscosity and 
magnetic fields only important in the former component. The highly asymmetric bar potential induces 
differential radial motions. While the gas component, being highly dissipative, suffers from the 
gravitational torque of the non-axisymmetric mass component,  the stars are mainly affected by orbital 
mixing. Furthermore, studies of the gas-phase abundances provide with present-day snap-shots of the interstellar 
medium abundance. On the other hand, the study of stellar abundances provide archeological clues as 
to the formation and evolution of the bar.  Therefore, the study of stellar and gas metallicities are 
of crucial importance to interpret the processes dominating galaxy evolution. It is expected that, 
if gas is injected into the interstellar medium (ISM) from stellar outputs, it would generally be more metal-rich  than 
if it has an external origin. 

Deriving the disk radial distribution of stellar abundances from spectroscopic measurements is difficult
due to the low surface brightness of the disk and the contamination of the emission lines from ionised gas. 
Few works have tried to overcome these difficulties and 
have derived the radial distribution for a small number of galaxies 
(e.g. Yoachim \& Dalcanton 2008; MacArthur, Gonz\'alez \& Courteau 2009; S\'anchez-Bl\'azquez et al. 2011). 
The stellar metallicity gradients along the bars of the 20 early-type galaxies presented here were carried 
out in previous works (P\'erez et al. 2009, hereinafter, Paper I) and provided interesting 
results regarding the radial distribution of the stellar parameters in the bar region. 

As for the nebular gas abundance distribution, for most spiral galaxies, both barred and unbarred, 
negative radial metallicity gradients along  the disk have been obtained  for the gas component 
(e.g. Bresolin et al. 2009). These gradients seem to be shallower in the case of barred galaxies than in 
unbarred galaxies, and larger for late-types than for early ones (Pagel \& Edmunds 1981; Alloin et al. 1981; 
Vila-Costas and Edmunds 1992; Zaritsky et al. 1994). The issue is not closed yet, so work is being done to investigate the variation of 
the gradient with radius (Bresolin et al. 2009 obtain a higher gradient in the central part of M83, 
and flatter in the outer disk, beyond 1.2 R$_{25}$) and also in azimuth (Balser et al. 2011, for the Milky Way).
Simulations explain the  shallower gradient value in the outer part than in the inner side of a galaxy 
based on  radial mixing processes associated to a strong bar (e.g. Considere et al. 2000; 
Zahid \& Bresolin 2011; Friedli et al. 1994; Friedli 1999). In this work, we will focus on the analysis of the 
gas metallicity, in the bar region of the galaxies presented in Papers I and II. 

Only a few works have considered both, nebular and stellar metallicities. A pioneering paper in this 
study is that of Storchi- Bergmann et al. (1994). They found a good correlation between stellar and 
gas metallicity investigating through the correlation between the oxygen abundance in the nebular 
component, and the absorption-line equivalent width W$({\rm C IV} \lambda 1550)$ (as a tracer of stellar metallicity) in a sample of 44 star-forming galaxies. 
They concluded that galaxies with lower values of metallicity appear to experience instantaneous 
outbreaks of star formation, but those with  higher metallicity are probably experimenting an ongoing 
stellar formation. Annibali et al. (2010) estimated gas abundance gradients in  early-type galaxies, and 
they compared these results with their  stellar metallicities (Annibali et al. 2007), concluding that the gas 
metallicity tends to be lower than the stellar one, beeing this effect higher for higher metallicities.

In this paper, we present an analysis of the radial oxygen abundance distribution in the bulge and 
bar region of  a sample of 
20 early type barred galaxies, with and without nuclear activity. The stellar component  of this sample 
has been studied in Papers I and II. In Sec. 2 observations and data reduction are described. 
The method to analyse the spectral lines to estimate their flux is presented in Sec. 3. In Sec. 4
we show the diagnostic diagrams and derive the abundance distributions with different methods.
A comparison of the nebular and stellar metallicity gradients is also presented in this section.
The results are discussed in Sec. 5. 
  
\section{Observations and data reduction}
   \label{observations}

We obtained long-slit spectra along the bar major axis\footnote{The bar position angles
were derived using the Digital Sky Survey (DSS) images.} for our sample  
of 20 barred galaxies. 
The   observations were performed in two different runs, with the double  
beam spectrograph  at Siding Spring Observatory (Australia,  
hereinafter, run1) and with the IDS spectrograph at the Isaac Newton  
Telescope (La Palma, Spain, hereinafter, run2).
The observations were described in detail in Paper I. To summarise,  
spectra for seven galaxies were obtained in the first run, covering a  
wavelength range from 3892-5815\AA\ and 5390-7314\AA\, with a spectral resolution of FHWM$\sim$2.2\AA. In spite of this overlap between both ranges, relative calibration is not possible because the
transmission of the dichroic gets as low as 50\%, decreasing the S/N relation, and we have no bright lines in this region. In the second run, spectra for 13 galaxies were  
obtained, covering a wavelength range of 3020-6665\AA\ with a spectral  
resolution of $\sim$3\AA~(FWHM), but the unvignetted range is 3600-6300\AA.

The seeing size during these observations
range from 0.6$''$ to 1.2$''$ at the INT and from 1$''$ to 1.5$''$ at SSO.
The characteristics of the sample are described in  Paper I and II. 
For convenience, Table~\ref{sample.tab} shows the sample main properties.

Atmospheric dispersion can have important
effects in spectrophotometry when the slit position angle differs from the
parallactic angle during the observations. This is not so critical in our
observations as we observed extended objects (diffuse emission or emission
knots with sizes typically a few arcseconds larger than our slit width).

In any case, to be in the safe side, we kept the target airmass below 1.6
during the observations for ensuring an atmospheric differential refraction
between [OII]3727\AA\ and [OIII]5007\AA\ smaller than 1.5" (Filippenko 1982). 
Differential atmospheric refraction between any other pair
or lines involved in the empirical metallicity calibrations used in the
paper are always smaller than the differential atmospheric refraction
between [OII] 3727\AA\ and [OIII]5007\AA.

For a few galaxies in which the difference between the bar position angle
and the parallactic angle was greater than $\approx$ 40 deg, we tried to
keep the airmass below  1.3 (which implies a differential atmospheric
refraction between  [OII]3727\AA\ and [OIII]5007\AA\ of $\approx$ 1").

The reduction of the two runs was carried out with the package  
{\tt REDUCEME} (Cardiel 1999)\nocite{cardiel} and {\tt IRAF} tasks.  Standard  
data reduction procedures (flat-fielding, cosmic ray removal,  
wavelength calibration and sky subtraction) were performed.  
Error images were created at the beginning of the reduction and were  
processed in parallel with the science images. For details about the  
reduction steps, see Paper I.

\begin{table*}
\begin{center}
\caption{General properties of the galaxy sample\label{sample.tab}.}
\begin{tabular}{l r l c l l r r r l l l}     
\hline\hline
Object            & Type & Bar class& Nuclear  
type & Inner morph.&$B$& V$_{max,gas}$ &  
$i$& Bar  & Bulge\\
 & & & & & &  & & size & radius & Run\\
   & & & & & & (km\,s$^{-1}$) &(deg) &(arcsec) & (arcsec) & \\
 (1)  &(2) &(3) &(4) &(5) &(6) & (7) &(8) &(9) & (10) & (11)\\
\hline
NGC~1169  & SABb    &3$^{a}$          & (--)             &    
--         &12.35        & 259.1 $\pm$ 7.3              & 57.1 & 29 & 5 & run2\\
NGC~1358           & SAB(R)0    &    --       &Sy2 (Sy2)           
&   --         &13.19        & 136.1 $\pm$ 10.6              & 62.8 & -- & 5 &run2\\
NGC~1433  &(R)SB(rs)ab&4$^{b}$&LINER (Sy2) &Double--bar$^{a}$&10.81&  
85.1 $\pm$ 2.4 & 68.1 & -- & 5 & run1\\
NGC~1530  &SBb&6$^{b}$&(--)&--&12.50&169.1$\pm$3.5&58.3 & 69 & 15 & run2\\
NGC~1832  &SB(r)bc&2$^{d}$&(--)&--&12.50&129.9$\pm$2.0& 71.8 & -- & 7 & run2\\
NGC~2217        &(R)SB(rs)0/a&--&LINER (LINER?)&Double--bar$^{b} 
$&11.36& 183.4$\pm$ 9.2 & 30.7 & -- & 5 & run1\\
NGC~2273  &SB(r)a&2$^{c}$&Sy2 (Sy2)&--&12.62&192.2$\pm$5.5&57.3 & 21 & 5 & run2\\
NGC~2523           &SBbc&--&(--)&--&12.64 &211.4 $\pm$10.9 &61.3 & --& 10 & run2\\
NGC~2665         &(R)SB(r)a&--& T (--)&--&12.47& 130.9$\pm$7.1&32.8 & -- & 5 & run1\\
NGC~2681  &(R)SAB(rs)0/a&1$^{c}$&(LINER)&Triple--bar$^{c} 
$&11.15&87.5$\pm$6.7&15.9 & 23 & 8 & run2\\
NGC~2859   &(R)SB(r)0~\^&1$^{c}$&(Sy)&Double--bar$^{d}$&11.86&  
238.5$\pm$13.3& 33.0 & 48 & 7 & run2 \\
NGC~2935          & (R)SAB(s)b  & --&T (--) &-- &12.26 &188.3$\pm 
$2.0 &42.7 &-- & 7 & run1\\
NGC~2950           &(R)SB(r)0\^~0&--&(--)&Double--bar$^{e} 
$&11.93& --& 62.0 & 44 & 10 & run2\\
NGC~2962           &(R)SAB(rs)0&--&(--)&Double--bar$^{c} 
$&12.91&202.9$\pm$9.9& 72.7 & 45 & 5 & run2\\
NGC~3081  &(R)SAB(r)0/a&3$^{c}$&Sy2 (Sy2)&Double--bar$^{d}$&12.89  
&99.9$\pm$4.0&60.1 & 35 & 5 & run1\\
NGC~4245    &SB(r)0/a&2$^{c}$&(--)&--&12.33&113.5$\pm$5.4&56.1 & 59 & 6 & run2\\
NGC~4314   &SB(rs)a&3$^{a}$&LINER (LINER)&Double--bar$^{c}$&11.42&253.3$ 
\pm$24.6&16.2 & 92 & 8 & run2\\
NGC~4394    &(R)SB(r)b&3$^{d}$&(LINER)&--&11.59& 212.5$\pm 
$16.0&20.0 & 56 & 5 & run2\\
NGC~4643  &SB(rs)0/a&3$^{c}$&LINER (LINER)&--&11.68&171.4$\pm$7.2&42.9 & 67 &10 & run1 \\
NGC~5101  &(R)SB(r)0/a&2$^{d}$&LINER (--)&--&11.59&195.7$\pm$9.0&23.2 &-- & 7 & run1\\
\hline
\end{tabular}
\end{center}
{\footnotesize
$(1)$ Galaxy identification name.
$(2)$ Galaxy morphological classification.
$(3)$ Bar class derived from:
$^a$ the $K-band$ light distribution, Block et al. (2001);
$^b$ the $K-band$ light distribution, Block et al. (2004);
$^c$ the $K-band$ light distribution, Buta et al. (2006);
$^d$ the $H-band$ light distribution, Laurikainen et al. (2004).
$(4)$ Nuclear type, from this paper when possible. Between ``()'' is the previous classification carried out by Veron-Cetty \& Veron (2006). $T$ is transition object. ``--'' means that no classification as active galaxy has been found in the literature.
$(5)$ Inner morphology, from
$^a$ Buta (1986)
$^b$ Jungwiert et al. (1997)
$^c$ Erwin (2004)
$^d$ Wozniak et al. (1995).
$(6)$, $(7)$ \& $(8)$ Apparent total B-magnitude, maximum rotational velocity corrected for inclination and
galaxy inclination, from the Hyperleda galaxy catalog. 
$(9)$ Bar semi-major axis, form Paper II.
$(10)$ Bulge size, from Paper II.
$(11)$ run1 are galaxies observed at the SSO and run2 those observed at the INT.} 
\end{table*}

\section{Analysis}
  \label{analysis}

We have observed 20 early type barred galaxies. Nebular emission has been detected in 19 of them. In NGC~2950, we detect no emission at all, and in NGC~4245 only at a galactocentric radius of 6 arcsec. In NGC~1832 and NGC~1530 we resolve knots of star formation (SF hereafter) at radius R$\le$ 0.25 R$_{25}$ ($R_{25}$ being the length of the projected major axis of a galaxy at the isophotal level 25 mag/arcsec$^2$ in the B-band). For the other 16 galaxies, we observe non-resolved emission extending at least along the bulge radius, i.e. $R \le 5''-15''$ depending on the galaxy (Table I). For seven galaxies (NGC~1433, NGC~1530, NGC~1832, NGC~2665, NGC~2935, NGC~3081 and NGC~4314) we have
also found knots of gas emission at larger radii, corresponding to one or
several crowded \hii\  regions.

To compare gas and stellar abundances we extract the spectra along the radius using the  
same binnings than in Paper I. These binnings were chosen to ensure errors lower than 15\% in most of the Lick/IDS indices.  We also add apertures at larger radius
to include those knots of star formation found in 7 of our galaxies (see above).

\begin{figure}
\centering
\includegraphics[width=\columnwidth]{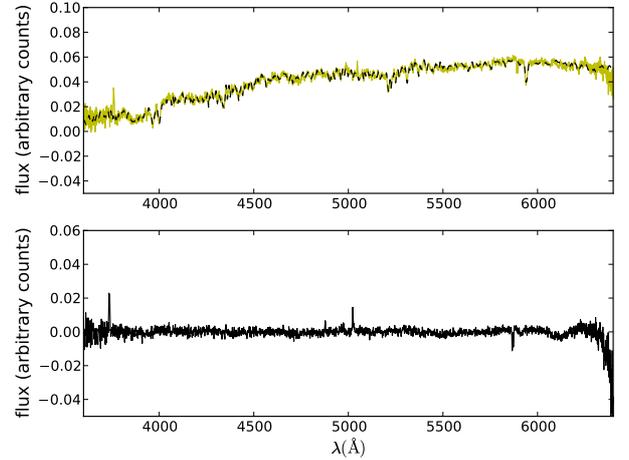}
\caption{Central spectrum for NGC~1169 (from run2). The top panel shows the observed spectrum (yellow line) and the fitted template (black line). Bottom panel shows the residuals from the fit. Oxygen and \hb\ lines are clearly visible.}
\label{espec_1169}%
\end{figure}

\begin{figure}
\centering
\includegraphics[width=\columnwidth]{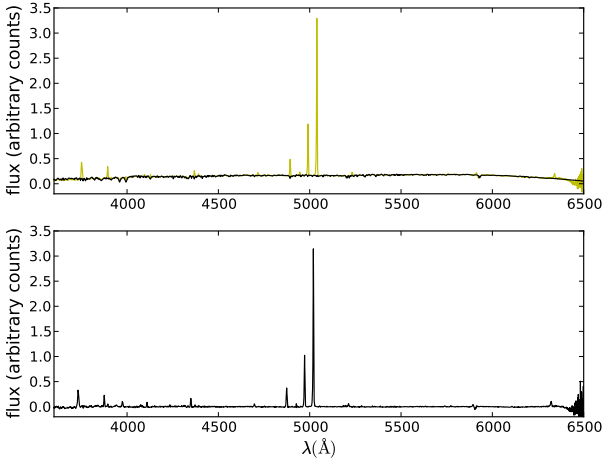}
\caption{Central spectrum for NGC~2273 (run2). Panels are the same as in Fig.~\ref{espec_1169}. In this spectrum it can be also identified the [\neiii]$\lambda$3868 line.}
\label{espec_2273}%
\end{figure}

\begin{figure}
\centering
\includegraphics[width=\columnwidth]{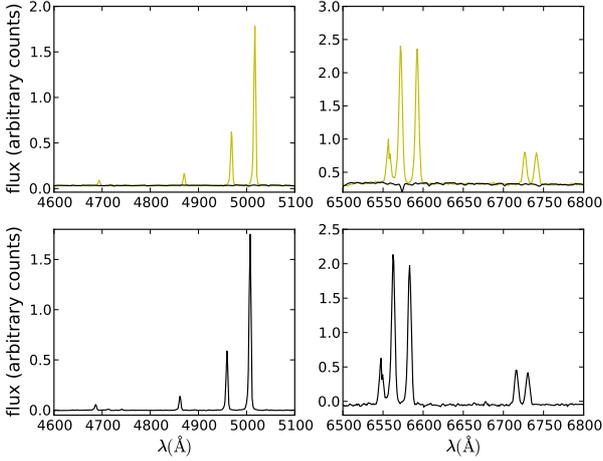}
\caption{Blue and red central spectra for NGC~3081 (run1). In the top pannels observed spectra (yellow lines) and templates (black lines) are shown. Bottom panels show the residuals of the fit.}
\label{espec_3081}%
\end{figure}

The gas emission line fluxes are measured from the spectra after subtracting 
stellar templates.  These stellar templates are
obtained using {\tt GANDALF} (Sarzi et al.\ 2005) and
the set of stellar population models by Vazdekis et al.\ (2010) using the MILES  library (S\'
anchez-Bl\'azquez et al.\ 2006). After the stellar template
subtraction, the spectra contains only the emission from the gas, with no
contribution from the underlying stellar population. As representative examples, we show in Figs.~\ref{espec_1169},~\ref{espec_2273} and ~\ref{espec_3081} the central spectrum for three galaxies, two of run2 (one of them with weak emission and the other with stronger emission lines) and one of run1. They are shown before and after the template subtraction.

The emission line fluxes are measured using the {\tt SPLOT } task in
IRAF, by integrating the line intensity over a locally fitted continuum
level. After marking two continuum points at both sides of the line, the flux is obtained taking into 
account the area, without a previous fitting to a Gaussian or to any other assumed profile. Therefore 
asymmetries due to dynamical processes are indirectly considered. 
The corresponding uncertainties are estimated adding in quadrature
statistical errors measured by {\tt SPLOT}, flatfielding errors and the
error of the flux calibration. Typical errors are of order $\sim$5\%  (Table~\ref{lineas} shows the errors for the central apertures).

The observed line ratios relative to \hb\   are afterwards corrected for
interstellar reddening, by determining the  reddening
coefficient c(\hb) through the expression:
\begin{equation}
{I(\lambda) \over {I(H\beta)}} = {F(\lambda) \over {F(H\beta)}}
10^{c(H\beta)\left[f(\lambda)-f(H\beta)\right]}
\end{equation}
where $f(\lambda)$ is the Seaton (1979) reddening curve parametrized by
Howarth (1984), and $I(\lambda)$  and $F(\lambda)$ are the intrinsic and
observed
emission line flux at wavelength $\lambda$, respectively.
We employ the Balmer lines detected within our spectral ranges. 
As explained in Sect. 2, both arms of the run1 observations have been treated separately for avoiding the introduction of an additional continuum scaling error, and therefore the \ha/\hb\ is not available.
The extinction constant has then been obtained from the average of the
extinction constant obtained from the fitting of all available Balmer line
ratios relative to  \hb, and the corresponding intrinsic or theoretical
values. We consider the intrinsic flux ratio from Hummer \& Storey (1987), assuming $T_e= 10^4$ K
and $N_e$= 100 cm$^{-3}$. 
For those cases in which there is a high dispersion between the values
for c(\hb) obtained from different lines ratios, we use the
extinction constant
obtained from H$\gamma$/\hb\  alone. We assign an internal extinction of 
zero to objects for which the observed Balmer decrement is lower than the theoretical values.
The derived reddening is negligible for most of the spectra. In the case of NGC 4245, 
we have obtained only \hb\ and no other Balmer line, but 
we have not considered this galaxy in the discussion. For NGC~1169, we observe only \hb\ except 
at 5 galactocentric radii. For 4 galaxies the observed Balmer decrement is lower
than the theoretical one, so we consider as zero the internal extinction. In the fifth case,
the obtained internal extinction is so low that differences between line fluxes considering or not
 this internal extinction are lower than the obtained errors. Therefore, we have assumed 
a negligible extinction in this galaxy.

Emission line strengths for the central aperture are shown in Table~\ref{lineas}. For NGC~4314 we show the emission line strengths for the innermost aperture (0.8 arcsec from the galaxy center) with detected emission. This central aperture corresponds to 0.75 arcsec for run1 galaxies and 0.4 arcsec for run2 galaxies. Values of [\oii], [\neiii] and [\oiii] are relative to \hb=100, and those of [\nii] and [\sii] to \ha=100.

For those galaxies presenting knots of gas emission, we also extract spectra
with apertures including all detected emission. In these cases, we correct from extinction using an iterative method to determine, simultaneously, the underlying stellar
absorption and
the interstellar reddening: we derive the  extinction coefficient
c(\hb) for an assumed equivalent width of the stellar absorption
EW$_{abs}$ (the same for all observed Balmer lines). The assumed
EW$_{abs}$ is varied until convergence between the c(\hb) 
obtained from all the Balmer line ratios in use.
The c(\hb) values are in the  range 0-2, 
and EW$_{abs}$ was found to be in the range 0.7-1.8.

\begin{table*}
\begin{center}
\caption{Emission Line Strengths for the central aperture. Emission line strengths for the central aperture of the galaxies, except for NGC~4314, which is at 0.8 arcsec from the galaxy center. The line fluxes of [\oii], [\neiii] and [\oiii] lines are relative to \hb\ of 100 and those of [\nii] and [\sii] lines are relative to \ha\ of 100.}             
\label{lineas}      
\centering          
\begin{tabular}{c c c c c c c c}     
\hline\hline       
Galaxy   & [\oii]        & [\neiii]     & [\oiii]       & [\oiii]          & [\nii]       &  [\sii]    &  [\sii]    \\ 
         & 3727          & 3868         & 4959          & 5007             & 6583         &  6717      &  6731      \\
\hline 
NGC~1169 & 525 $\pm$ 25  & ---          & ---           &  207 $\pm$ 10    & ---          &  ---       & ---         \\
NGC~1358 & 278 $\pm$ 13  & 65 $\pm$ 3   & 370 $\pm$ 18  & 1153 $\pm$ 55    & ---          &  ---       & ---         \\
NGC~1433 & ---           & ---          & 51  $\pm$ 2   &  155 $\pm$ 7     & 103 $\pm$  6 & 34 $\pm$ 2 & 28 $\pm$ 2   \\
NGC~1530 & 135 $\pm$ 40  & ---          & ---           &  23.5$\pm$ 1.5   & ---          &  ---       & ---         \\
NGC~1832 & 28.6$\pm$1.4  & ---          & 26.9$\pm$ 1.3 & 33.5 $\pm$ 1.6   & ---          &  ---       & ---          \\
NGC~2217 & ---           & ---          & 100 $\pm$ 5   &  306 $\pm$ 14    & 226 $\pm$ 13 & 131$\pm$ 11& 115 $\pm$ 10  \\
NGC~2273 & 168 $\pm$ 8   & 68 $\pm$ 3   & 237 $\pm$ 11  &  184 $\pm$  9    & ---          &  ---       & ---          \\
NGC~2523 & 211 $\pm$ 10  & ---          & 194 $\pm$ 9   &  306 $\pm$ 15    & ---          &  ---       & ---          \\
NGC~2665 & ---           & ---          &  23 $\pm$ 1   &   72 $\pm$  3    &  67 $\pm$  3 & 19 $\pm$ 1 & 20  $\pm$ 1  \\
NGC~2681 &  71 $\pm$ 3   & ---          & 105 $\pm$ 5   &  164 $\pm$  8    & ---          &  ---       & ---         \\
NGC~2859 & 421 $\pm$ 20  & ---          &  99 $\pm$ 5   &  178 $\pm$  8    & ---          &  ---       & ---           \\
NGC~2935 & ---           & ---          &  45 $\pm$ 2   &  128 $\pm$  6    &  71 $\pm$  3 & 24 $\pm$ 1 & 19 $\pm$ 1  \\
NGC~2950 & ---           & ---          & ---           & ---              & ---          & ---        & ---          \\
NGC~2962 & 399 $\pm$ 19  & ---          &  67 $\pm$ 3   &  174 $\pm$  8    & ---          &  ---       & ---          \\
NGC~3081 & ---           & 84 $\pm$ 4  & 414 $\pm$ 20  & 1242 $\pm$ 60    &  88 $\pm$  6 & 23 $\pm$ 1 & 21$\pm$ 2 \\
NGC~4245 & ---           & ---          & ---           & ---              & ---          & ---        & ---          \\
NGC~4314 &230 $\pm$ 11  & ---         &  93 $\pm$ 4   &  203 $\pm$ 10    & ---          & ---        & ---           \\
NGC~4394 & 289 $\pm$ 14  & ---          & 108 $\pm$ 5   &  205 $\pm$ 10    & ---          & ---        & ---          \\
NGC~4643 & ---           & ---          & 115 $\pm$ 5   &  73  $\pm$  3    &  98 $\pm$ 12 & 31 $\pm$ 5 & ---          \\  
NGC~5101 & ---           & ---          &  86 $\pm$ 4   &  114 $\pm$  5    & 189 $\pm$ 12 & 75$\pm$  7 &  59 $\pm$ 6  \\
\hline                  
\end{tabular}
\end{center}
{\footnotesize }
\end{table*}

\section{Results}
  \label{results}

\subsection{Star formation versus nuclear activity}

Once the line fluxes are measured and extinction corrected, we can represent diagnostic diagrams 
commonly used to distinguish between star formation (SF) regions and regions containing or affected by 
an active nucleus (AGN). However, classical diagnostic diagrams ($\log$([\oiii]/\hb) {\it versus} 
$\log$([\nii]/\ha) and $\log$([\oiii]/\hb) {\it versus} $\log$([\sii]/\ha)) 
(Baldwin et al. 1981, Veilleux \& Osterbrock 1987) can only be applied to run1, due to the limited 
spectral coverage of run2. For four galaxies, one from run1 and three from run2  we detect the 
[\neiii]$\lambda$3868 line. This line has been proposed as an empirical indicator of metallicity 
(Nagao et al. 2006) and also as a diagnostic to distinguish between starburst galaxies and 
AGNs (see discussion in P\'erez-Montero et al. 2007).

\begin{figure}
\centering
\includegraphics[width=\columnwidth]{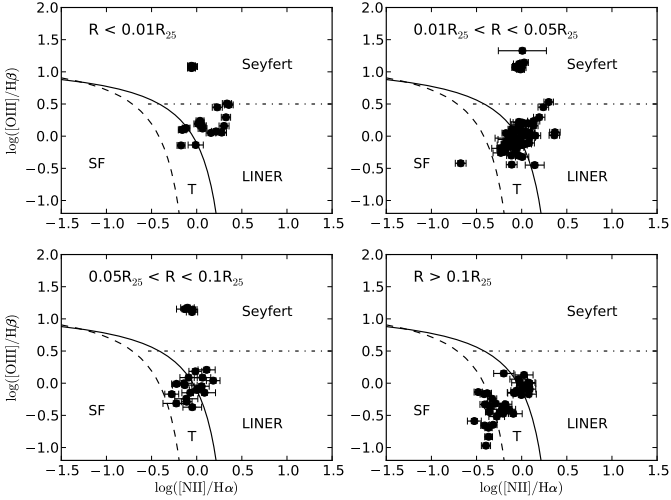}
\caption{Diagnostic diagram for the emission-line regions of run1 galaxies at different galactocentric 
distances (R) scaled with R$_{25}$. The solid line (Kewley et al. 2001) sets the star formation upper 
limit, the dashed line  (Kauffmann et al. 2003) the empirical division between star-forming galaxies 
(SF) and transition objects (T), and the horizontal line (Kewley et al. 2006) separates Seyferts and LINERs.}
\label{diag_todas_1}%
\end{figure}

\begin{figure}
\centering
\includegraphics[width=\columnwidth]{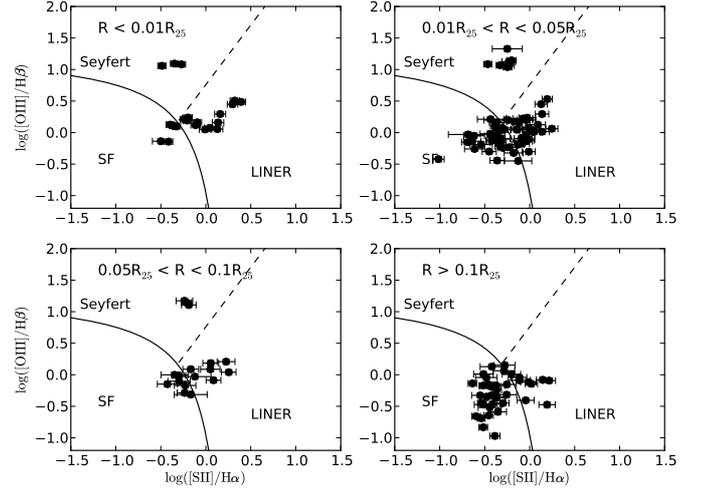}
\caption{Diagnostic diagram for the emission-line regions of run1 galaxies for different galactocentric 
distances (R) scaled with R$_{25}$. The solid line establishes the separation between star formation regions 
and AGNs, below the dashed one there are LINERs and above it Seyferts (Kewley et al. 2006).}
\label{diag_todas_2}%
\end{figure}

\begin{figure}
\centering
\includegraphics[width=\columnwidth]{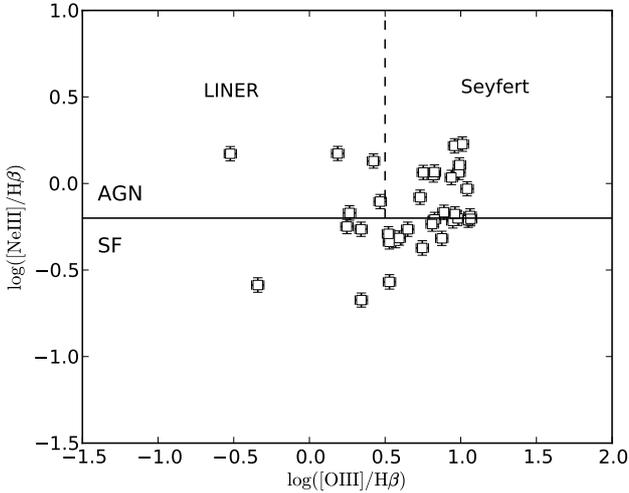}
\caption{Diagnostic diagram for those galaxies in which we have detected the [\neiii]$\lambda 3869$ line. 
The solid line indicates the limit between star formation regions and AGNs, and the dashed one between 
LINERs ([\oiii]5007/\hb $\le$ 0.5) and Seyferts 2 ([\oiii]5007/\hb $>$ 0.5) (Rola et al. 1997).}
\label{diag_todas_Ne}%
\end{figure}

Figures~\ref{diag_todas_1},~\ref{diag_todas_2} and ~\ref{diag_todas_Ne} show the diagnostic diagrams. 
Figures~\ref{diag_todas_1} and ~\ref{diag_todas_2} show these diagnostic diagrams at different galactocentric distances. 
In Fig.~\ref{diag_todas_1} we have plotted the theoretical curves of Kewley et al. (2001), 
which sets the maximun expected [\oiii]/\hb\ for a given [\nii]/\ha\ for a star forming region, those 
from Kauffmann et al. (2003) which sets the empirical division between star-forming galaxies and 
transition objects, and that of Kewley et al. (2006) separating Seyfert and LINERs galaxies. In  
Fig.~\ref{diag_todas_2}  we plot the relation between the line ratios $\log$([\sii]/\ha) and 
$\log$([\oiii]/\hb) for the galaxies of our sample at different radii. The theoretical models 
separating the values corresponding to ionization by star formation and AGN are also indicated. 
Rola et al. (1997) proposed an alternative diagnostic diagram based on the [\neiii]$\lambda$3869 line.
In Fig.~\ref{diag_todas_Ne}  this diagnostic diagram for galaxies in which we have detected this line is plotted, relating  $\log$([\neiii]/\hb) with $\log$([\oiii]/\hb).  The solid line indicates the limit between star formation regions and AGNs, and the dashed one between LINERs ([\oiii]5007/\hb $\le$ 0.5) and Seyferts 2 ([\oiii]5007/\hb $>$ 0.5) as explained in Rola et al. (1997).

Figures~\ref{diag_todas_1} and ~\ref{diag_todas_2} also show that at large radii there is a larger probability that the gas is ionized by star forming regions, while central emission is characterized by AGN-like line-ratios. 

Studying the diagnostic diagrams for individual galaxies, we confirm the classification carried out by  
Veron-Cetty \& Veron (2006) with the exception of NGC~5101 and NGC~1433.  According to the emission 
line in the central region, we reclassify NGC~5101 as a LINER even if it  was not classified in such a way in Veron-Cetty \& Veron (2006),  but in agreement with Moiseev (2001). The run1 galaxies that are not classified as Seyfert or LINER,  NGC~2665 and NGC~2935, are transition objects. In Table I we show the nuclear type obtained in this paper and in previous classification carried out by Veron-Cetty \& Veron (2006).

\begin{figure*}
   \includegraphics[width=20cm]{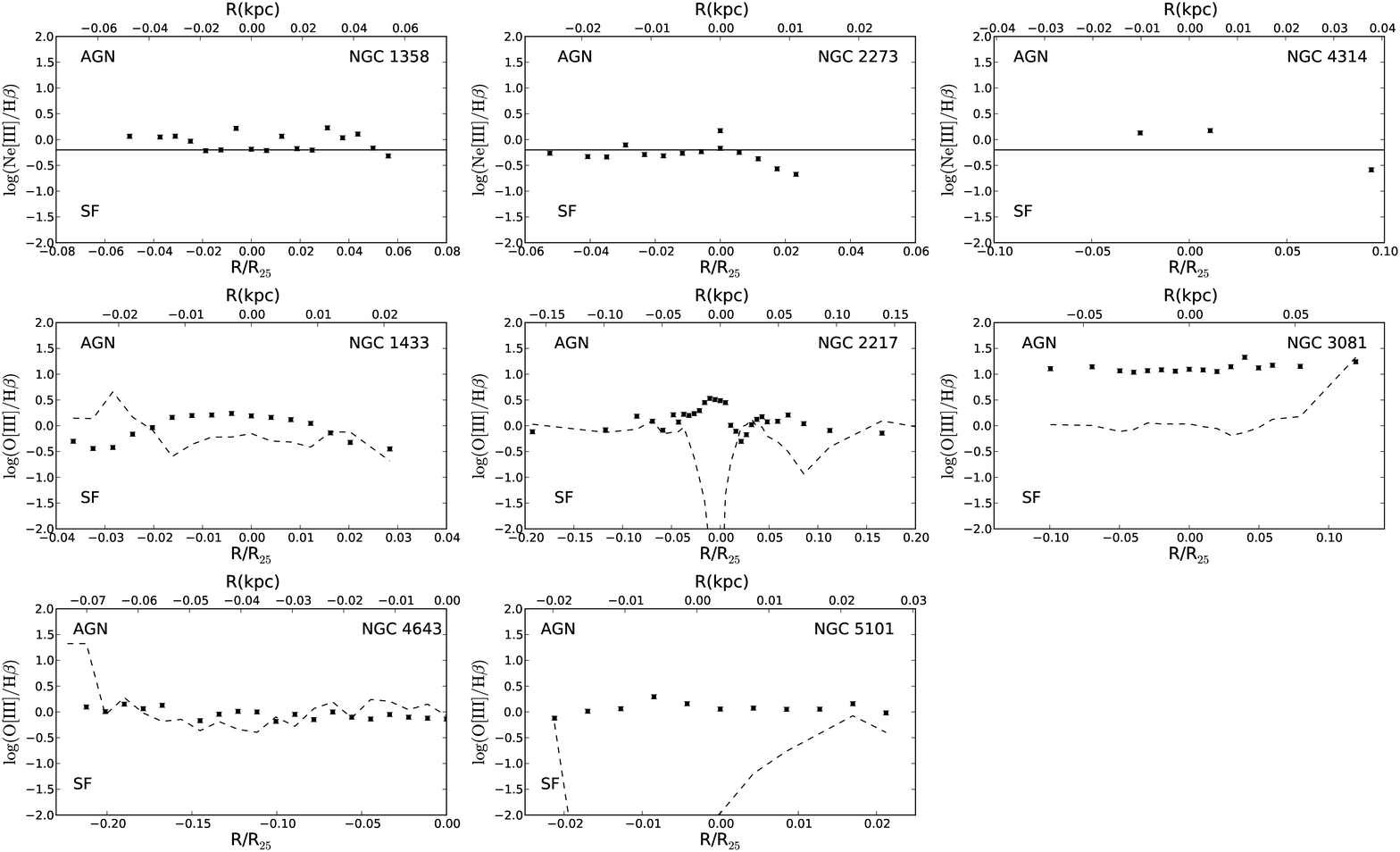}
      \caption{Radial distribution of $\log$([\oiii]/\hb) (for active galaxies of run1) and $\log$([\neiii]/\hb) 
      (for run2 galaxies with  [\neiii]$\lambda 3869$ line detected). The theoretical curves of 
      Kewley et al. (2001, for $\log$([\oiii]/\hb) and $\log$([\nii]/\ha)) and Rola et al. (1997, 
      for $\log$([\neiii]/\hb)) distinguishing AGN and SF are also plotted with dashed and solid lines, 
      respectively.}
         \label{actividad}%
   \end{figure*}

 Since we have radial information for the diagnostic diagrams, we have considered the question of the 
 AGN influence region by analyzing the radial behavior of the diagnostic ratios, to find the radius at which 
 star formation starts being the main photoionisation source. Fig.~\ref{actividad} shows the radial 
 distribution of $\log$([\oiii]/\hb) (or $\log$([\neiii]/\hb)). In the same figure we have considered the measured values of [\nii]/\ha. With these values and taking into account the theoretical curve of Kewley et al. (2001) (or Rola et al. 1997 for $\log$([\neiii]/\hb)) we obtain the limiting values to consider star formation or nuclear activity. They are indicated with a dashed line. At a given galactocentric radius, if the observational points are 
 above it, a star forming region is not the main photoionization component at that radius. There are differences 
 among galaxies. The AGN or LINER is the dominant ionization 
 mechanism up to different radii for the different galaxies: For example, for NGC~2273 it is just AGN ionized at the lowest radius point. For NGC~2217 the influence region of AGN extends to $\sim$ 0.1 kpc.

\subsection{Oxygen abundance and gradients}

The measurement of nebular gas abundances presents several difficulties. The {\it direct method} to 
calculate oxygen abundances implies the use of  ratios between temperature sensitive lines, such 
as [\oiii]$\lambda \lambda$4959,5007/$\lambda$4363 (Osterbrock 1989). Unfortunately, when the metallicity 
increases, the electronic temperature decreases and the auroral lines (e.g. [\oiii]$\lambda$4363) become too weak 
to be measured in most extragalactic objects and, therefore, we have to find empirical, semiempirical and theoretical calibrations to 
calculate abundances. Some relations between the gas metallicity and line flux ratios of strong emission 
lines have been found. A drawback of these indicators is that they have been basically developed for 
low-metallicity \hii\ regions, where the [\oiii]$\lambda$4363 line is easily observed. However, a considerable 
effort to study high metallicity \hii\ regions has also been made 
(e.g. Castellanos, Diaz \& Terlevich 2002, Bresolin 2007), since these are the most abundant in 
early spiral galaxies and in the central regions of late type spiral galaxies  (Vila-Costas \& Edmunds 1992). 
Although semiempirical methods provide better results in statistical studies than  individual values 
(P\'erez-Montero et al. 2007), we use these methods to estimate the oxygen abundance variations with 
galactocentric radius.
Systematic discrepancies among different methods have been found (see, for instance, Kewley \& Ellison 2008; L\'opez-S\'anchez \& Esteban 2010, and references therein), so we use several methods to study the dependence of the results on the method chosen.

An additional problem is that 11 out of our 20 galaxies have been classified as LINERs or Seyferts. The mechanism of ionization in the case of LINERs is still not well understood, so there are no methods to estimate abundances in the central parts of these galaxies. Some methods exist in the case of Seyferts, but due to spectral coverage we can only apply one of them, and in some particular cases. We also apply this method in LINERs, which may be intermediate between AGN excitation or photoionization due to star formation.

\subsubsection{Nebular abundances: Errors in their determination and comparison between different methods}

We use different methods to estimate the nebular abundances for the two different runs, as the spectral 
range covered in both data sets is different. The brightest lines measured, apart from the Balmer 
lines (H$\gamma$, \hb, \ha) are [\oiii]$\lambda \lambda$4959,~5007, [\nii]$\lambda\lambda$6548,~6583, 
[\sii]$\lambda\lambda$6717,~6731 for run1 galaxies, and [\oii]$\lambda$3727 and 
[\oiii]$\lambda \lambda$4959,~5007 for run2 galaxies. At some distances from the center 
[\oiii]$\lambda$4959 is not detected, in which case we assume [\oiii]$\lambda\lambda$4959,5007=1.337[\oiii]$\lambda$5007 (Oey \& Kennicutt 1993).

The abundance indicators used in this work are the following:
\begin{enumerate}
\item {\it R23}: Based on this parameter ($R_ {23}= \left([OII]\lambda 3727 + [OIII]\lambda \lambda 4959,5007]\right)/ H\beta$), we use two calibrations: 
\begin{itemize}
\item M91: The theoretical calibration of McGaugh (1991), which takes into account the influence of 
the ionization parameter to determine the chemical abundance. 
\item Z94: It is an analytical calibration obtained by Zaristky et al. (1994) as the average of three 
previous  $R_{23}$ calibrations (Dopita \& Evans 1986; McCall et al. 1985; Edmunds \& Pagel 1984). It is 
only valid for high metallicity objects with $\log($O/H)+12 $>$ 8.35.
\end{itemize}
\item {\it N2}: It uses $\left({[NII]\lambda 6583} \over {H\alpha}\right)$ (Pettini \& Pagel 2004) 
with the advantages of not being double valued with oxygen abundance, and that lines are very near each 
other, minimizing the dependence on the reddening correction. 
\item {\it O3N2}: It depends on $\left({[OIII]\lambda 5007/H\beta} \over {[NII]\lambda 6583/H\alpha}\right)$ 
(Pettini \& Pagel 2004), with the advantage of being a monotonic function of the oxygen abundance. 
However, in this case, the pairs of lines ([\oiii]$\lambda$5007, \hb) and ([\nii]$\lambda$6583, \ha) are 
detected in separated spectra, which can introduce errors caused by a missmatch of the continuum fluxes in 
the red and blue (run1) ranges. It will be discussed at the end of this Section.
\item {\it O2Ne3}: P\'erez-Montero et al. (2007) discussed the relation between [\neiii]$\lambda$3869 and [\oii], and define the parameter $O2Ne3={{I([OII]\lambda 3727)+15.371([NeIII]\lambda 3869)} \over {I(H\beta)}}$. Taking into account this parameter, the abundance is calculated with the expression of $M91$. This line is more intense in AGNs due that, in these objects, Ne$^{+2}$ is less suppressed than in star formation regions, as explained in Rola et al. (1997).

\item {\it SB1}: It is calibrated for active galaxies (Storchi-Bergmann et al. 1998) and has two 
different  calibrations. It is valid for Seyfert but not so accurate for LINERs. We use the calibration involving 
the [\oiii], [\nii], \ha\ and \hb\ lines because of our observed spectral range. It depends on the gas density, 
being valid only for 100$<$N$<$10000 (with N the gas density in cm$^{-3}$) and $8.4 < 12+\log(O/H) < 9.4$.
\item {\it KD02}: Method proposed by Kewley \& Dopita (2002) which consists on an iterative scheme 
involving the ionization parameter and $R_{23}$. 
\item {\it P05}: Pilyugin \& Thuan (2005) improved the calibration of Pilyugin (2000, 2001) with a more 
extended (in metallicity) set of \hii\ regions. The lines needed to measure this indicator 
are [\oii]$\lambda\lambda$~3727,3729, 
[\oiii]$\lambda\lambda$~4959,5007 and \hb.
\end{enumerate}

{\it R$_{23}$}, {\it O2Ne3}, {\it KD03}, {\it P05} methods are applied to galaxies observed in run2 and 
{\it N2}, {\it O3N2}, {\it SB1} to those observed in run1. 

Several of the indices described above show two branches in their calibration, {\it i.e.},  
there are two possible metallicity values  for a given index. This makes  necessary to 
have further criteria to distinguish between the upper and the lower branches. We adopt the high-metallicity 
values on the basis of (a) the empirical diagrams of Nagao et al. (2006) (though not for active galaxies): With a sample of 50000 spectra they have proposed a method, based in the monotonic metallicity dependence of [\oiii]$\lambda$ 5007/[\oii]$\lambda$ 3727, to break the $R_{23}$ degeneracy. We have made use of their diagrams to determine the branch.
(b) the Kobulnicky et al. (1999) conclusion about galaxies in the local Universe (i.e. objects more luminous 
than $M_B \sim -18$ have metallicities higher than 8.3, and the whole  sample lies 
within this range), (c)  morphologically, all the galaxies are of early type and Oey \& Kennicutt (1993) 
found that the average metallicity value for \hii\ regions in Sa and Sb galaxies is 8.97, clearly in the 
upper region, and (d) in NGC~1530 we obtain in this upper branch values similar to those found by 
M\'arquez et al. (2002) using the [\nii]/H$\alpha$ ratio. 
An special case is that of NGC~1358: the {\it R$_{23}$} values calculated for this galaxy are 
in the limit of applicability of the methods based on this parameter. But with data of [\nii] and H$\alpha$ we have calculated abundance values supporting those obtained with  {\it R$_{23}$}.

The errors in the oxygen abundances have been estimated from error
propagation on the uncertainties in the relevant line fluxes and range between $\sim$0.01
and 0.10 dex (or 10-20\% in [O/H]). This is lower than
the intrinsic scatter associated to empirical calibrations of strong line
methods (e.g. 0.2 dex for R$_{23}$, Athey \& Bregman 2009).
We have carried out several tests to determine the uncertainties introduced in the different parts 
of the analysis. For that, abundances are estimated in two possible ways for each test of this analysis. In what follows, we describe the maximum difference found and the mean value of these differences in a galaxy, chosen as a characteristic example:\\
(a) As discussed in Sect. 3, in most of the spectra we have used stellar templates to substract 
the contribution from the underlying stellar population, being a second possibility to correct the line intensities 
with an iterative method. The differences found in gaseous abundance between both methods are lower than 0.1-0.2 dex. As an example, for NGC~1433 the mean values of these differences are 0.04 $\pm$ 0.06 dex for {\it N2}, 0.05 $\pm$ 0.03 dex for {\it O3N2} and 0.04 $\pm$ 0.02 dex for {\it SB1}\\
(b) Considering  either an equivalent width due to Balmer absorption of 
2\AA\  to correct the stellar absorption, a typical value used in the literature (e.g. Zaritsky et al. 1994), or using the 
iterative method to estimate it, introduces differences lower than 0.1-0.2 dex. For example, in NGC~1530 the mean values of these differences are 0.04 $\pm$ 0.04 dex ({\it R$_{23}$}), 0.05 $\pm$ 0.04 dex ({\it Z94}), 0.03 $\pm$ 0.02 dex ({\it KD03}) and 0.13 $\pm$ 0.09 dex ({\it P05})\\
(c) The error introduced if we consider  an I(\ha)/I(\hb) value of either 2.85 or 3.1 is as low as 0.05 dex. Annibaldi et al. 2010 also showed that, in any case, considering \ha/\hb\ = 2.85  does not  
significantly change the results. For NGC~3081, mean differences are 0.0004 $\pm$ 0.0008 dex ({\it N2}), 0.003 $\pm$ 0.004 dex ({\it O3N2}) and 0.08 $\pm$ 0.09 dex ({\it SB1})\\
We next show that differences among the different metallicity calibrations are larger 
than these errors.

\begin{figure*}
   \includegraphics[width=\textwidth]{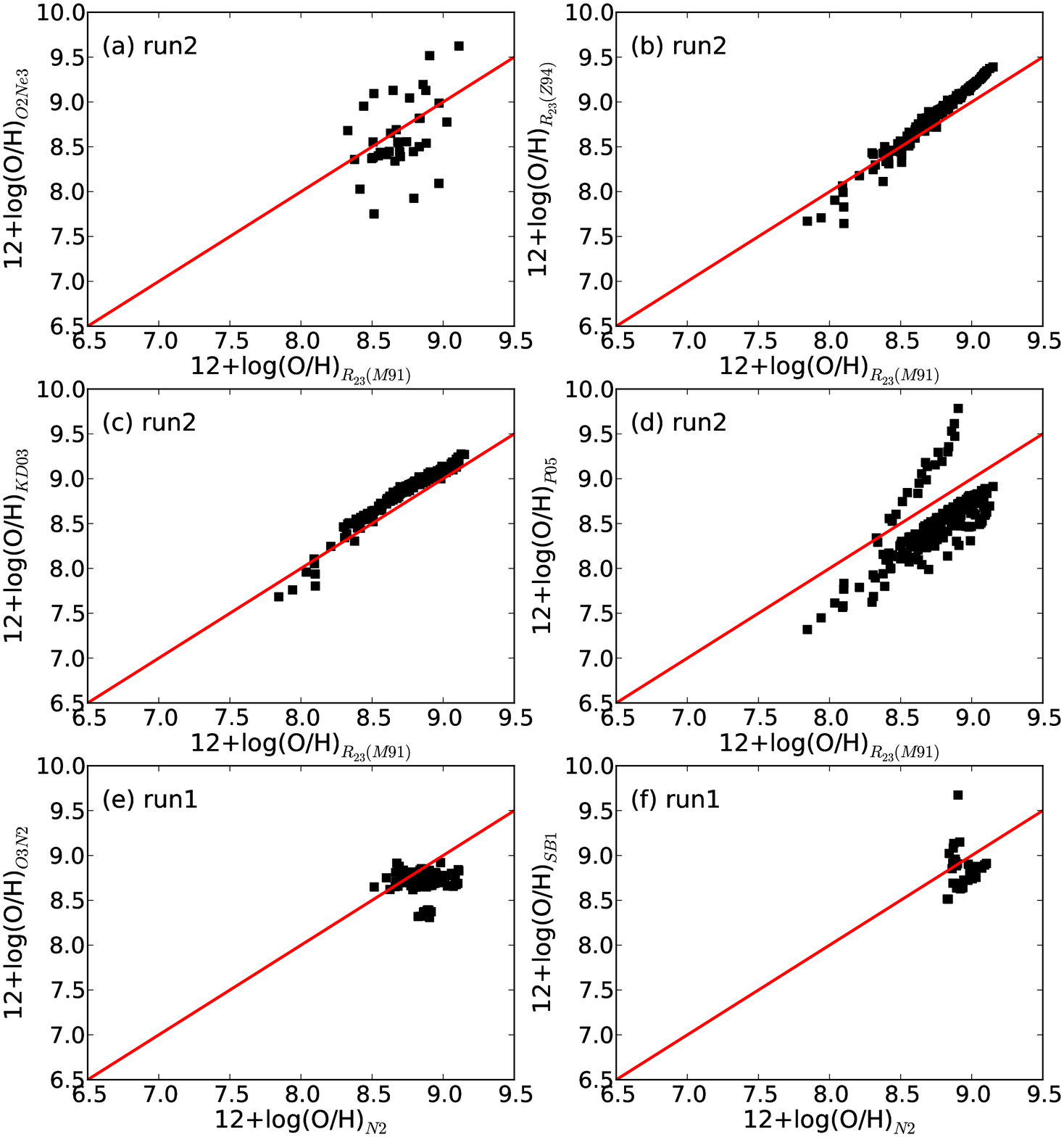}
   \caption{Relation between different calibrations for obtaining oxygen abundances. 
   In (a), (b), (c) and (d) we compare the methods used for run2 galaxies with the {\it R$_{23}$} calibration by McGaugh (M91). 
In (e) and (f) the comparison is for run1 galaxies, between {\it N2} and {\it O3N2} or {\it SB1}. The solid line in each
 plot shows the agreement between the different methods.}
              \label{comresG}%
    \end{figure*}

We compare the results from the different methods in Fig.~\ref{comresG}, taking {\it N2} as reference 
for run1 galaxies, and {\it R$_{23}$}(M91) for run2 galaxies. {\it O2Ne3} can only be calculated for 
three galaxies; its behavior is the same  as the {\it R$_{23}$} (M91) but with a larger dispersion. 
{\it  R$_{23}$} (Z94) is valid for  $12+\log{(O/H)} > 8.35$ (Zaritsky et al. 1994).  
For $12+\log{(O/H)}$ values between 8.35 and 8.7, {\it R$_{23}$}, with the calibrations of Z94 and M91, 
provide similar metallicity values, but from 8.7 upwards, the Z94 value becomes higher than the M91 one. 
In {\it KD03} the behavior is the same  as in {\it R$_{23}$} (M91), but {\it KD03} gives values slightly 
above those of {\it R$_{23}$} (M91)  from 8.2 upwards. {\it P05} values are systematically 0.3-0.5 dex 
lower than those of {\it R$_{23}$} (M91) with the exception of NGC~1358. Bresolin et al. (2009) showed that, 
for the external parts of the disk of M83, the metallicity  with [\nii]/[\oii] is closer to the actual values, 
being 0.4 dex lower than those obtained with {\it R$_{23}$}. Therefore,  it might be expected that the metallicities
using the {\it P05} method  are more realistic or, at least, closer to the direct metallicity determinations. 
Kennicutt et al. (2003) found systematic differences up to 0.5 dex between $T_e$ and {\it R$_{23}$} abundances. 
Pilyugin \& Thuan (2005) noted that the determination of oxygen abundances using {\it P05} agrees with 
$T_e$ abundance to within 0.1 dex, also for high-metallicity ones, which is our case. 

For run1 galaxies, the largest differences between {\it O3N2} and the {\it N2} index are 0.3-0.4 dex, 
except for NGC~3081, in which case the difference is around 0.5 dex. Denicol\'o et al. (2002) emphasized the 
importance of the {\it N2} method, showing that the {\it O3N2} calibration did not improve the 
results. {\it SB1} is a method calibrated for AGN; though eleven of these galaxies have been classified as 
active, this method has only been used in the cases that meet the above conditions of density and metallicity, and only to run1 galaxies because of the lines involved.

\begin{figure*}
   \centering
   \includegraphics[width=0.9\textwidth]{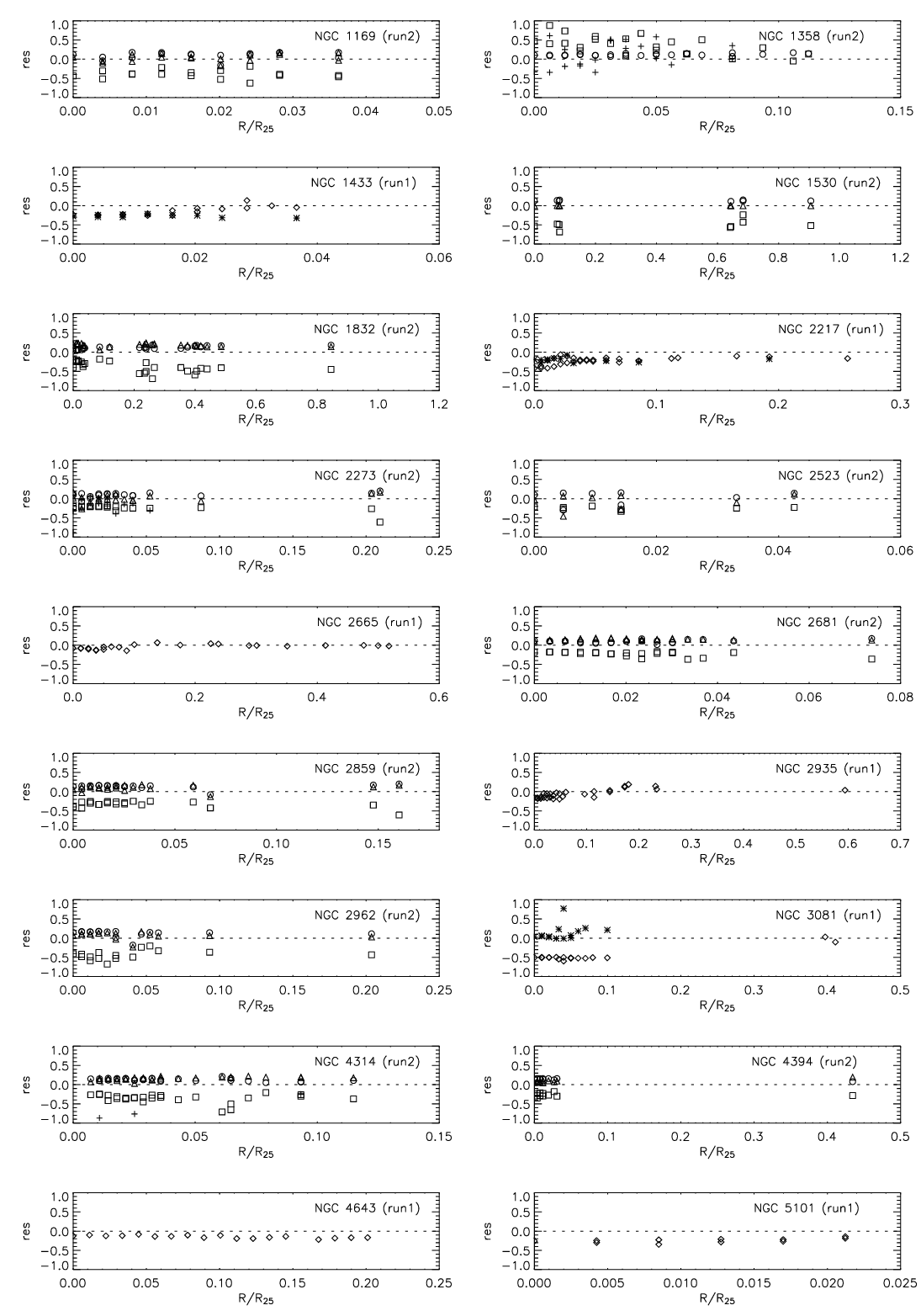}
   \caption{Differences between the oxygen abundances obtained with the different methods versus the
   galactocentric radius for each galaxy. $R_{25}$ has been taken from HyperLeda. 
   Residuals are with respect to {\it R$_{23}$} (M91) for run2 galaxies and to {\it N2} for run1 galaxies. 
   Crosses are oxygen abundances obtained from the {\it O2Ne3} calibration, triangles from {\it R$_{23}$} (Z94), circles from {\it KD03}, squares from {\it P05}, diamons from {\it O3N2} and asteriks from {\it SB1}.}
              \label{res_TODAS}%
\end{figure*}

It is interesting to compare the radial dependence of the values obtained with the different methods, 
to study the effect of nuclear activity or the differences due to  physical differencies in 
the nuclear and disk \hii\ regions properties. \hii\ characteristics can influence the abundance trends obtained using different methods 
in different ways  (Kennicutt et al. 1989). The systematic differences will be  considered in the comparison with 
stellar metallicities. Figure~\ref{res_TODAS} shows the residual values  of the different abundance 
calculation methods for all the galaxies with respect to {\it R$_{23}$} (M91) for run2 galaxies and 
{\it N2} for run1 galaxies.

As already noted, {\it P05} abundances are always below {\it R$_{23}$} (M91), the difference being around 
0.5 dex, without changes at different radii (with the expection of the special case NGC~1358 as we have mentioned). 
For {\it KD03}, the largest differences are found in the central part of the Seyfert galaxy NGC~1358, 
being much smaller beyond 0.08 R/R$_{25}$, the activity zone in Fig.~\ref{actividad}. But this is not the 
general behavior for Seyferts (see NGC~2273), for which we find similar values to those calculated with 
{\it R$_{23}$} calibrations by M91 and Z94.  We find a larger dispersion for {\it O2Ne3} than for the 
other methods, without a systematic trend with radius, being the largest deviation for galaxies with 
less bins (e.g. NGC~4314). Between {\it O3N2} and {\it N2} the differences are small in most galaxies, 
less than 0.2-0.3 dex. But in two of them, NGC~3081 and NGC~2217, these differences are higher 
(0.4-0.5 dex), and with the peculiarity that become smaller (again 0.1-0.2 dex) for larger galactocentric distances.

As above mentioned, 11 galaxies have been classified as Seyferts or LINERs. All methods used here to derive metallicity, with the exception of {\it SB1}, have been derived for star-forming galaxies. We can only use the specific calibration method  {\it SB1} in three out of our twenty galaxies: 
NGC~1433, NGC~2217 and NGC~3081, even only one of these, i.e. NGC~3081, is clearly a Sy2. Annibali et al. (2010) compared oxygen abundances estimated from a photoionization model based on {\it R$_{23}$} (Kobulnicky et al. 1999) with {\it SB1}. The values obtained from the {\it SB1} method are on average $\sim$0.04 dex larger than those using the {\it R$_{23}$} calibration, but the differences between both metallicity values in individual regions can be as large as 0.3 dex. Keeping in mind this result we have opted to show the values obtained with all the methods for comparison. This can be specially useful in the case of LINERs, as for them the mechanism of ionization is not known. Therefore we have calculated the abundance values with {\it SB1} when possible. For the active galaxies for which this method cannot be used
we only give metallicities obtained from methods calibrated for star forming regions, and we warn again the reader about the validity of these values in the AGN dominated area.

 \begin{figure*}
   \includegraphics[width=1.1\textwidth]{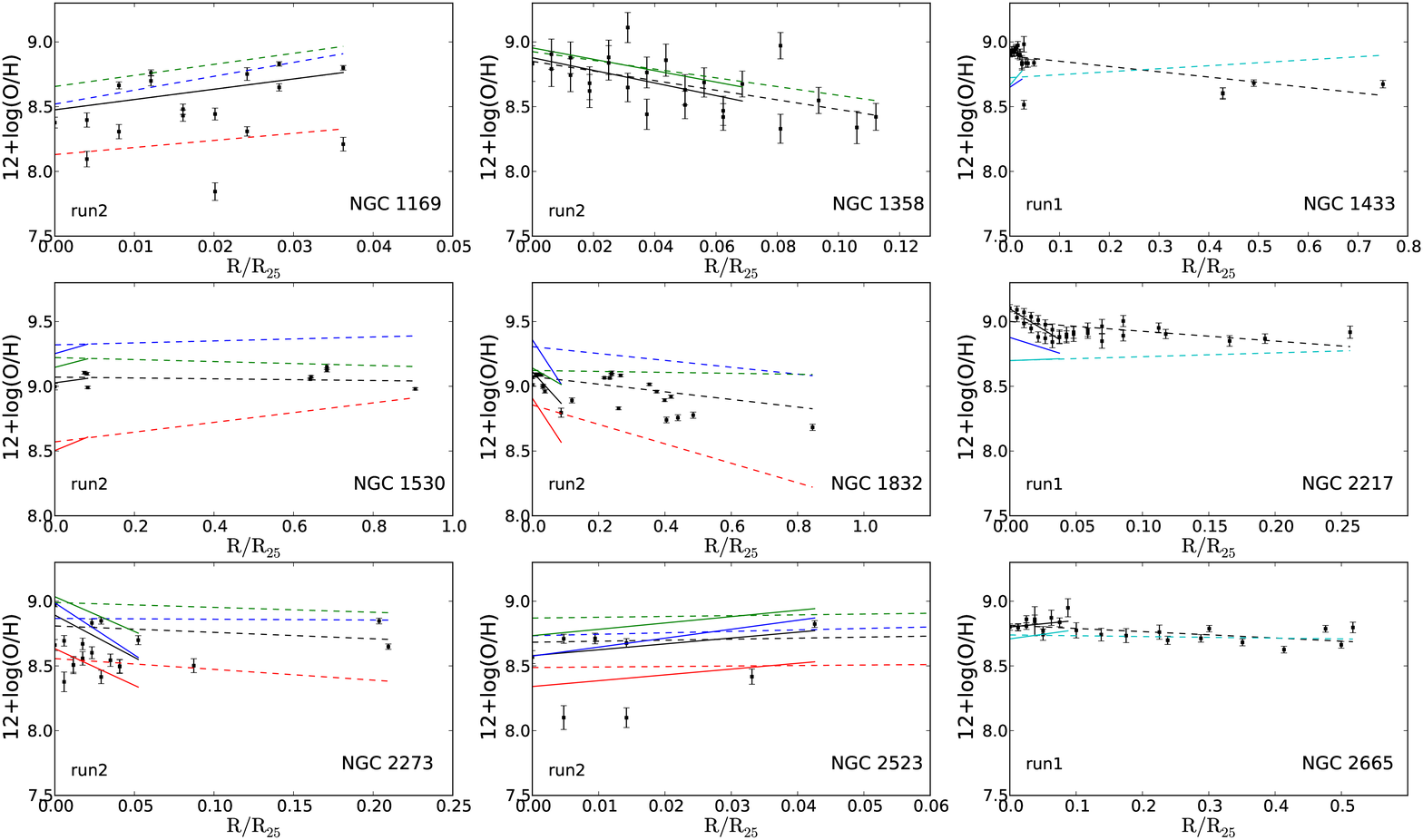}
   \caption{12+log(O/H) obtained with {\it O3N2} for run1, and with {\it R$_{23}$} (M91) for run2, versus R/R$_{25}$ in each galaxy. Dashed lines indicate the fit for all measured values, and solid lines show the best fits within the bulge region. For run1 galaxies, black lines indicate the {\it N2}, cyan lines the {\it O3N2} fit and blue line the {\it SB1} fit. For run2 galaxies black lines show the {\it R$_{23}$} (M91) fit, blue lines the {\it Z94} fit, green lines the {\it KD03} fit and red lines the {\it P05} fit.}
              \label{ajuste1}%
    \end{figure*}

 \begin{figure*}
  \setcounter{figure}{9}
   \includegraphics[width=1.1\textwidth]{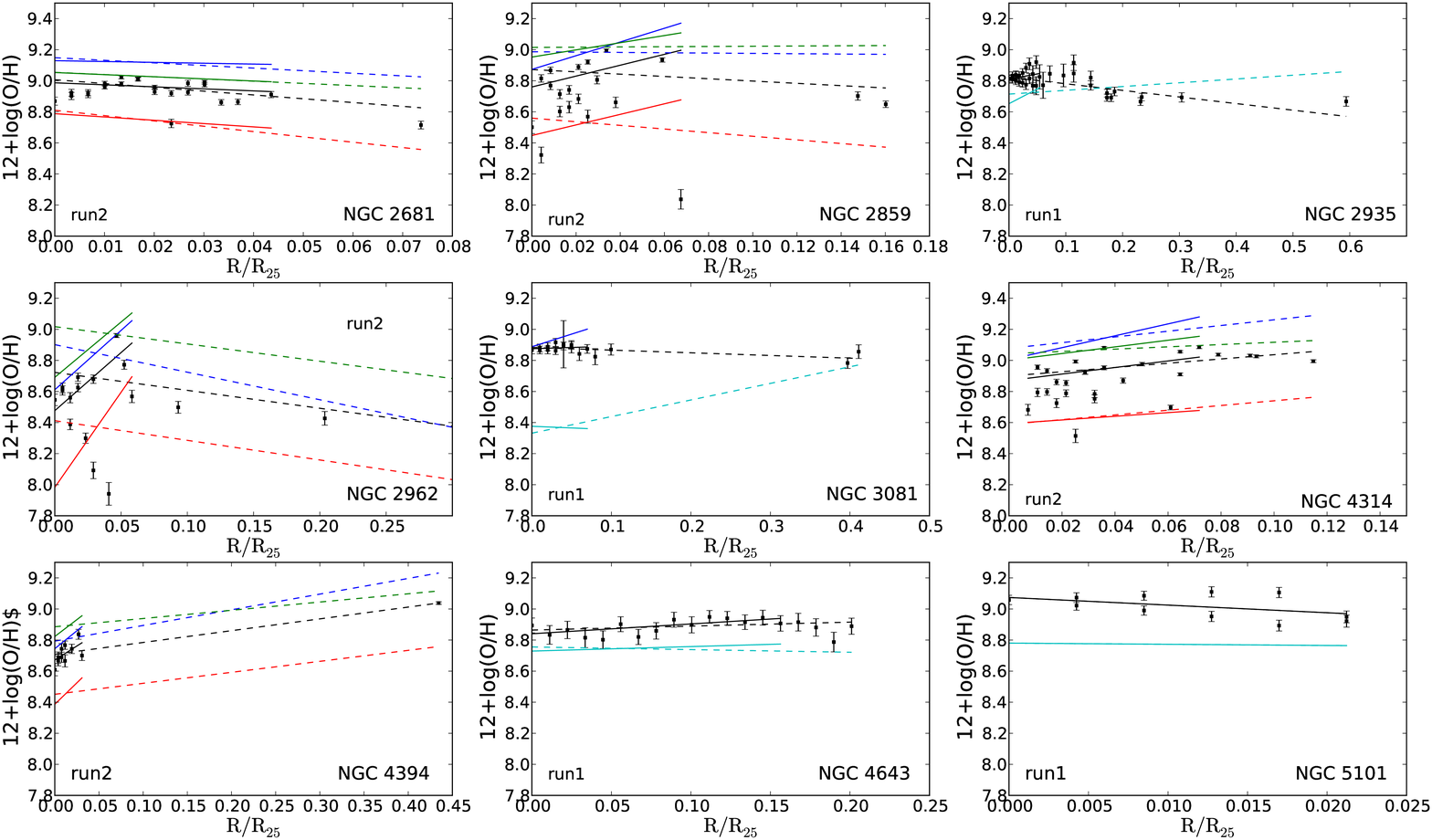}
   \caption{Continued}
    \end{figure*}

\begin{table*}
\begin{center}
\caption{Central abundances, for run2 galaxies.}             
\label{gradientes_INT_R_1}             
\begin{tabular}{l c c c r r}     
\hline\hline       
Galaxy    & 12+log(O/H)$^{(1)}$ & 12+log(O/H)$^{(2)}$ & 12+log(O/H)$^{(3)}$ &  R$_{25}^{(4)}$ & AGN$^{(5)}$\\ 
                    & dex      & dex & dex &  arcsec & arcsec  \\
\hline 
NGC~1169 & 8.47  $\pm$ 0.10 &  8.47  $\pm$ 0.01 &   &  99.34 & \\
NGC~1358$\dagger$   & 8.78  $\pm$ 0.07 &  8.85  $\pm$ 0.04 &  8.71   $\pm$ 0.12 & 64.14 & 3.2\\
NGC~1530 & 8.90  $\pm$ 0.02 &  9.071 $\pm$ 0.006&  & 54.59 &\\
NGC~1832 & 9.09  $\pm$ 0.01 &  9.075 $\pm$ 0.003&  & 73.64 &\\ 
NGC~2273$\dagger$  & 8.57  $\pm$ 0.05 &  8.809 $\pm$ 0.008&   8.828  $\pm$ 0.009& 68.72 & 0.4\\
NGC~2523 & 8.48  $\pm$ 0.05 &  8.50  $\pm$ 0.10 &  & 84.55 &\\
NGC~2681$\dagger$ & 8.92  $\pm$ 0.02 &  9.007 $\pm$ 0.007 &  9.12  $\pm$ 0.02 & 119.43 & 3$^*$\\ 
NGC~2859$\dagger$ & 8.66  $\pm$ 0.03 &  8.872 $\pm$ 0.007 &  9.05  $\pm$  0.01 & 94.87 & 3$^*$\\
NGC~2962 & 8.58  $\pm$ 0.03 &  8.72  $\pm$ 0.01 & & 68.72 &\\
NGC~4314$\dagger$& 8.81  $\pm$ 0.02 &  8.900 $\pm$ 0.004 &  8.965 $\pm$ 0.008 & 111.46 & 3.3\\
NGC~4394$\dagger$ & 8.69  $\pm$ 0.03 &  8.708 $\pm$ 0.009&  8.67  $\pm$  0.02 & 104.02 & 3$^*$\\
\hline                  
\end{tabular}
\end{center}
{\footnotesize 
$\dagger$Active galaxies.\\
$^{(1)}$Central oxygen abundance in the gas, estimated with an aperture of 1.2 arcsec, and calculated with the {\it R$_{23}$} (M91) method.\\
$^{(2)}$Central oxygen abundance in the gas, as a result of the fit, calculated with the {\it R$_{23}$} (M91) method.\\
$^{(3)}$Central oxygen abundance in the gas, as a result of the fit outside the region of AGN influence, calculated with the {\it R$_{23}$} (M91) method.\\
$^{(4)}$R$_{25}$ is from the Hyperleda galaxy catalogue.\\
$^{(5)}$AGN influence region, in arcsec, obtained from Fig.~\ref{actividad} except those marked with *, which values are obtained from V\'eron-Cetty \& V\'eron (2006) and references therein.}
\end{table*}

\begin{table*}
\begin{center}
\caption{Oxygen abundance gradients with all used methods and the central abundance, for run2 galaxies.}             
\label{gradientes_INT_R_2}             
\begin{tabular}{l c c c c c}     
\hline\hline       
Galaxy   & grad-{\it $R_{23}$}(M91)    & grad-{\it O2Ne3}    & grad-{\it $R_{23}$}(Z94)     & grad-{\it KD03}     & grad- {\it P05}   \\ 
         & dex/arcsec         &  dex/arcsec        &  dex/arcsec       & dex/arcsec       & dex/arcsec   \\
\hline 
NGC~1169 & 0.080  $\pm$ 0.007 & ---                & 0.108  $\pm$ 0.015  & 0.086   $\pm$ 0.012  & 0.055   $\pm$ 0.013 \\
NGC~1358$\dagger$ & $-$0.058 $\pm$ 0.017 & 0.04  $\pm$ 0.03   & 0.11   $\pm$ 0.01   & $-$0.05   $\pm$ 0.01   & $-$0.12   $\pm$ 0.05  \\
               & $-$0.03 $\pm$ 0.02   & 0.06  $\pm$ 0.08   & 0.05  $\pm$ 0.03    & $-$0.03  $\pm$ 0.02    & $-$0.08  $\pm$ 0.10      \\
NGC~1530 & $-$0.0006$\pm$ 0.0002& ---                & 0.0014 $\pm$ 0.0002 & $-$0.0014 $\pm$ 0.0003 & 0.0069  $\pm$ 0.0002\\
NGC~1832 &$-$0.0040 $\pm$ 0.0002& ---                &$-$0.0036 $\pm$ 0.0003 &$-$0.0005  $\pm$ 0.0002 & $-$0.0102 $\pm$ 0.0002 \\ 
NGC~2273$\dagger$ &$-$0.0072 $\pm$ 0.0013&$-$0.01  $\pm$ 0.02   &$-$0.001  $\pm$ 0.002  &$-$0.0055  $\pm$ 0.0015 & $-$0.012  $\pm$ 0.002 \\
               & $-$0.009 $\pm$  0.001& $-$0.07  $\pm$ 0.02  & $-$0.004 $\pm$ 0.002  & $-$0.006  $\pm$  0.001 & $-$0.014  $\pm$ 0.002   \\
NGC~2523 & 0.0093 $\pm$ 0.0004& ---                & 0.0133 $\pm$ 0.0008 & 0.0073  $\pm$ 0.0005 &  0.0045 $\pm$ 0.0007\\
NGC~2681$\dagger$ &$-$0.020  $\pm$ 0.002 & ---          &$-$0.014  $\pm$ 0.004  &$-$0.012   $\pm$ 0.001  & $-$0.029  $\pm$ 0.003 \\ 
               &$-$0.046  $\pm$ 0.005 & ---          &$-$0.051  $\pm$ 0.009  & 0.010   $\pm$ 0.004  & $-$0.074  $\pm$ 0.008      \\
NGC~2859$\dagger$ &$-$0.0078 $\pm$ 0.0013& ---          &$-$0.001  $\pm$ 0.003  &0.000    $\pm$ 0.002  & $-$0.012  $\pm$ 0.003 \\
               &$-$0.027  $\pm$ 0.002 & ---          &$-$0.028  $\pm$ 0.004  &$-$0.020   $\pm$ 0.003  & $-$0.038  $\pm$ 0.003  \\ 
NGC~2962 &$-$0.017  $\pm$ 0.002 & ---                &$-$0.026  $\pm$ 0.004  &$-$0.016   $\pm$ 0.004  & $-$0.018  $\pm$ 0.003 \\
NGC~4314$\dagger$ & 0.0123 $\pm$ 0.0007& 0.111  $\pm$ 0.014 & 0.0165 $\pm$ 0.0014 & 0.0066  $\pm$ 0.0007 &  0.0136 $\pm$ 0.0012\\
               & 0.005  $\pm$ 0.001 &  ---               & 0.0016 $\pm$ 0.0019 & 0.0015  $\pm$ 0.0012 &  0.0046 $\pm$ 0.0016      \\
NGC~4394$\dagger$ & 0.0073 $\pm$ 0.0003& ---                & 0.0097 $\pm$ 0.0005 & 0.0051  $\pm$ 0.0003 &  0.0068 $\pm$ 0.0004\\
               & 0.0080 $\pm$ 0.0008& ---                & 0.0104 $\pm$ 0.0014 & 0.006   $\pm$  0.001 &  0.0084 $\pm$ 0.0012  \\
\hline                  
\end{tabular}
\end{center}
{\footnotesize 
$\dagger$Active galaxies (Seyferts and LINERs): The first row shows values calculated with all points, and the second one those calculated in the region outside the AGN influence.}
\end{table*}

\begin{table*}
\begin{center}
\caption{Oxygen abundance gradients with all the used methods, for run1 galaxies. Units are dex/arcsec.}             
\label{gradientes_SSO_R}      
\centering          
\begin{tabular}{l c c c c c c r r}     
\hline\hline       
Galaxy   & grad-{\it N2}    &  grad-{\it O3N2}       & grad-{\it SB1}   & 12+log(O/H)$^{(1)}$ & 12+log(O/H)$^{(2)}$ & R$_{25}^{(3)}$ & AGN\\ 
         & dex/arcsec       &  dex/arcsec            &  dex/arcsec      & dex  & dex & arcsec & arcsec\\
\hline 
NGC 1433$\dagger$ & $-$0.0022 $\pm$ 0.0002 &   0.0012 $\pm$ 0.0002 & 0.01  $\pm$ 0.01  & 8.68 $\pm$ 0.03 & 8.724 $\pm$ 0.008 & 184.98 & \\
               & $-$0.0014 $\pm$ 0.0002 &   0.0009 $\pm$ 0.0002 &                   &                 & 8.77  $\pm$ 0.01  &      & 3.7 \\ 
NGC 2217$\dagger$ & $-$0.0054 $\pm$ 0.0009 &   0.0021 $\pm$ 0.0008 & $-$0.023 $\pm$ 0.007 & 8.674 $\pm$ 0.03 & 8.699 $\pm$ 0.009 & 140.32 &\\
               & 0.0004  $\pm$ 0.0025 &  $-$0.0002 $\pm$ 0.0009 &                    &                  & 8.76  $\pm$ 0.03  &    &16.8 \\
NGC 2665 & $-$0.0041 $\pm$ 0.0006 &  $-$0.0010 $\pm$ 0.0006 &   ---                & 8.718 $\pm$ 0.03 & 8.739 $\pm$ 0.009 & 59.86 &\\
NGC 2935 & $-$0.0034 $\pm$ 0.0003 &   0.0019 $\pm$ 0.0004 &   ---                & 8.649 $\pm$ 0.02 & 8.714 $\pm$ 0.007 & 125.06 &\\ 
NGC 3081$\dagger$ & $-$0.0023 $\pm$ 0.0010 &   0.0142 $\pm$ 0.0009 &  0.024  $\pm$ 0.012  & 8.369 $\pm$ 0.02 & 8.332 $\pm$ 0.009 & 75.36 & \\
               & $-$0.0025 $\pm$ 0.0021 &  0.0187  $\pm$ 0.0018 &                      &                  & 8.22  $\pm$ 0.04  &      & 7.5\\
NGC 4643$\dagger$ &  0.004  $\pm$ 0.003  &  $-$0.003  $\pm$ 0.002  &   ---                & 8.751 $\pm$ 0.06 & 8.756 $\pm$ 0.019 & 67.16 &13.2\\
NGC 5101$\dagger$ & $-$0.028  $\pm$ 0.008  &  $-$0.004  $\pm$ 0.008  &   ---                & 8.784 $\pm$ 0.03 & 8.780 $\pm$ 0.019 & 176.66 &3.5\\  
\hline                  
\end{tabular}
\end{center}
{\footnotesize $^{(1)}$Central oxygen abundance in the gas, estimated with an aperture of 1.2 arcsec, and calculated with the {\it O3N2} method.\\
$^{(2)}$Central oxygen abundance in the gas, as a result of the fit, calculated with the {\it O3N2} method.\\
$^{(3)}$R$_{25}$ is from the Hyperleda galaxy catalogue.\\
$^{(5)}$AGN influence region, in arcsec, obtained from Fig.~\ref{actividad}.\\
$\dagger$Same as Table~\ref{gradientes_INT_R_2}}
\end{table*}

\subsubsection {Gas oxygen abundance in barred galaxies}

Figure~\ref{ajuste1} shows the oxygen  abundances obtained with {\it R$_{23}$} (M91) or {\it O3N2} {\it versus} radii for all galaxies. Linear fits, performed in the bar region and in the bulge region taking into account the errors derived from the line fluxes, are also plotted for the different methods.

Tables~\ref{gradientes_INT_R_1} and ~\ref{gradientes_SSO_R} show, respectively for run2 and run1, the value of oxygen abundances considering an aperture of 1.2 arcsec (averaging abundances inside 1.2 arcsec), its central value from the extrapolation down to R=0 of the linear fit (only for {\it R$_{23}$}(M91) and {\it O3N2}), and the  central value from the extrapolation down to R=0 excluding the influence region of the AGN. Tables~\ref{gradientes_INT_R_2} and ~\ref{gradientes_SSO_R} show gradients for all  used methods (in units of 
 dex/arcsec) taking into account all the nebular emission in the bar region and their errors. Galaxies classified as Seyferts or LINERs are marked with a dagger. For the galaxies for which we have enough points, we have two rows in these tables: The first one shows values obtained in the bar region, and the second one those obtained excluding the AGN influence region determined in the previous section. These influence regions have been determined from Fig.~\ref{actividad} except for NGC~2681, NGC~2859 and NGC~4394 for which we have considered the sizes reported by  V\'eron-Cetty \& V\'eron (2006) and references therein. The numerical values are shown in Tables~\ref{gradientes_INT_R_1} and \ref{gradientes_SSO_R}. 

The seeing size could introduce errors in estimating the gradients. 
To study this effect we have produced artificial spectra at each pixel
position. We have then convolved them with Gaussians of different width, covering the seeing values of these observations. 
This was made using the {\tt IRAF} {\tt GAUSS} task. We 
then generated three spectra. The largest differences 
in abundances calculated for these three spectra are lower than 0.2, 
implying differences in the gradient of $\sim$ 5 $\%$.

We analyze the results using  the indices {\it R$_{23}$}(M91) and {\it O3N2}.  For run2 galaxies differences are nearly constant at all radii, in such a way that gradients maintain their sign in nearly all methods. In run1 galaxies, there are changes with galactocentric distance as above mentioned. In particular, the three galaxies for which we have calculated SB1 coincide with those with the most different values of abundances in the central region, being smaller for {\it O3N2} than for {\it N2}. This difference does not keep at larger radii, so the gradient sign is opposite for both methods. The oxygen abundances in a 1.2 arcsec aperture are in the range 8.4-9.1 dex.

If we consider all the detected emission in the bar region at a 3-$\sigma$ significance, for the star formation galaxies, we detect a positive gradient for three galaxies (NGC~1169, NGC~2523 and NGC~2935) and a negative gradient is observed for two galaxies (NGC~1832 and NGC~2962). For LINERs, we find a positive gradient for three galaxies (NGC~1433, NGC~4314 and NGC~4394) and none of them has a negative gradient. Considering the region without influence of the possible AGN in these galaxies, the sign of the gradients does not vary. In the case of Seyferts we find negative gradients in four galaxies when we consider the bar region, and excluding the AGN influence region, three of them maintain the gradient sign. Only one of the Seyfert galaxies shows a positive gradient, which is not affected by the AGN region.

In most of the disks of spiral galaxies, both barred or unbarred, negative metallicity gradients have 
been obtained (e.g. Bresolin et al. 2009, Garnett \& Shields 1987, etc.). This gradient seems to be lower for 
barred galaxies than for unbarred ones, and higher for late type galaxies than for early type 
(Hidalgo-G\'amez et al. 2011, Vila-Costas \& Edmunds 1992, Zaritsky et al. 1994). All our galaxies
are barred and early type; so, based on previous studies, the gradient should be shallow. 
We find gradients in the range [-3.7,1.4] (normalized to R$_{25}$) with the exception of NGC~1169, which has a very high dispersion in the abundance values. In Sect. 5 we compare these values with those found in the literature.

With respect to the bulge region, in Tables~\ref{gradientes_INT_bulbo} and ~\ref{gradientes_SSO_bulbo} are shown the obtained gradients for all used method, and the central oxygen abundance (only with $R_{23}$ {\it (M91)} or {\it O3N2} method). In these tables we have marked the active galaxies  but the gradients are fitted including the AGN regions. NGC~3081 is the only Seyfert galaxy for which we have been able to calculate abundances with the {\it SB1} method. For two galaxies (NGC~1169 and NGC~5101) all the measured nebular emission in the bar position is in the bulge zone where the scatter is larger than in the bar region and, therefore, errors are also larger. We find the same sign for the slopes in most of the star-forming galaxies. They are in general steeper than those considering all the bar region.

\begin{table*}
\begin{center}
\caption{Metallicity gradients in the bulge regions with all the used methods for run2 galaxies,  including AGN regions.}         
\label{gradientes_INT_bulbo}              
\begin{tabular}{l c c c c c c c}     
\hline\hline       
Galaxy   & grad-{\it $R_{23}$(M91)} & 12+log(O/H)$^{(1)}$ & grad-{\it O2Ne3}   & grad-{\it $R_{23}$(Z94)}     & grad-{\it KD03}     & grad-{\it P05}   & order$^{(2)}$ \\ 
         & dex/arcsec    &      dex    &  dex/arcsec      &  dex/arcsec       & dex/arcsec                & dex/arcsec      &       \\
\hline 
NGC 1169 & 0.080  $\pm$ 0.007  & 8.47  $\pm$ 0.01  & ---               & 0.108  $\pm$ 0.015   & 0.086  $\pm$ 0.012    & 0.055  $\pm$ 0.013   & 1\\
NGC 1358$\dagger$ &$-$0.08 $\pm$ 0.02   & 8.88  $\pm$ 0.05  & 0.05  $\pm$ 0.04  & 0.16   $\pm$ 0.03    & $-$0.069  $\pm$ 0.019 & $-$0.15  $\pm$ 0.10  & 2\\
NGC 1530 & 0.008  $\pm$ 0.006  & 9.03  $\pm$ 0.02  & ---               & 0.016  $\pm$ 0.013   & 0.015 $\pm$ 0.009     & 0.023 $\pm$ 0.006    & 3\\
NGC 1832 &$-$0.038$\pm$ 0.003  & 9.113 $\pm$ 0.007 & ---               &$-$0.053$\pm$ 0.005   &$-$0.020 $\pm$ 0.003   &$-$0.053 $\pm$ 0.004  & 4\\ 
NGC 2273$\dagger$ &$-$0.095$\pm$ 0.007  & 8.892 $\pm$ 0.014 & -0.01  $\pm$  0.02&$-$0.117$\pm$ 0.012   &$-$0.078  $\pm$ 0.007  &$-$0.08  $\pm$ 0.01   & 5\\
NGC 2523 & 0.054  $\pm$ 0.010  & 8.58  $\pm$ 0.02  & ---               & 0.082  $\pm$ 0.017   & 0.058  $\pm$ 0.011    & 0.053  $\pm$ 0.014   & 6\\
NGC 2681$\dagger$ &$-$0.011$\pm$ 0.003  & 8.987 $\pm$ 0.007 & --                &$-$0.005$\pm$ 0.005   &$-$0.011 $\pm$ 0.001   &$-$0.018  $\pm$ 0.004 & 7 \\  
NGC 2859$\dagger$ &  0.037 $\pm$ 0.003  & 8.762 $\pm$ 0.009 & ---               &0.046   $\pm$ 0.006   &0.024  $\pm$ 0.003     &0.036  $\pm$ 0.006    & 8\\ 
NGC 2962 & 0.109  $\pm$ 0.006  & 8.48  $\pm$ 0.01  & ---               &0.111   $\pm$ 0.012   &0.104  $\pm$ 0.009     & 0.178  $\pm$ 0.010   & 9\\
NGC 4314$\dagger$ & 0.0189 $\pm$ 0.0012 & 8.871 $\pm$ 0.005 & -0.11 $\pm$0.16   & 0.034  $\pm$ 0.002   & 0.0192 $\pm$ 0.0015   &0.0106 $\pm$ 0.0019   & 10\\
NGC 4394$\dagger$ & 0.035  $\pm$ 0.010  & 8.671 $\pm$ 0.017 & ---               & 0.044  $\pm$ 0.017   & 0.042 $\pm$ 0.011     & 0.053  $\pm$ 0.015   & 11\\
\hline                  
\end{tabular}
\end{center}
{\footnotesize $^{(1)}$Central oxygen abundance in the gas, as a result of the fit in the bulge region, calculated 
with the {\it R$_{23}$} (M91) method.\\
$^{(2)}$See Figs.~\ref{grad-gas-estre-barras} \& ~\ref{grad-gas-estre-bulbos}\\
$\dagger$Same as Table~\ref{gradientes_INT_R_2}}
\end{table*}

\begin{table*}
\begin{center}
\caption{Metallicity gradients in the bulge regions, with all the used methods, for run1 galaxies, including AGN regions.}             
\label{gradientes_SSO_bulbo}      
\centering          
\begin{tabular}{l c c c c c}     
\hline\hline       
Galaxy   & grad-{\it N2}        & 12+log(O/H)$^{(1)}$& grad-{\it O3N2}       & grad-{\it SB1}  & order$^{(2)}$\\ 
         & dex/arcsec        &      dex        &  dex/arcsec        &  dex/arcsec  &\\
\hline 
NGC 1433$\dagger$ & $-$0.001 $\pm$ 0.006   & 8.665 $\pm$ 0.017 & 0.024  $\pm$ 0.006    & 0.01  $\pm$ 0.01 & 12\\
NGC 2217$\dagger$ & $-$0.045 $\pm$ 0.006   & 8.698 $\pm$ 0.018 &0.003   $\pm$ 0.005    &$-$0.023  $\pm$ 0.012 & 13\\
NGC 2665 &  0.008   $\pm$ 0.007   & 8.709 $\pm$ 0.018 & 0.012  $\pm$ 0.006    &                      & 14  \\
NGC 2935 &  0.005   $\pm$ 0.004   & 8.653 $\pm$ 0.016 & 0.015  $\pm$ 0.004    &                      &15\\ 
NGC 3081$\dagger$ &  0.002   $\pm$ 0.005   & 8.376 $\pm$ 0.015 &$-$0.003$\pm$ 0.005    & 0.022    $\pm$ 0.015 & 16 \\
NGC 4643$\dagger$ &  0.009   $\pm$ 0.004   & 8.73  $\pm$ 0.02  & 0.004  $\pm$ 0.004    &                      & 17  \\
NGC 5101$\dagger$ & $-$0.028 $\pm$ 0.008   & 8.780 $\pm$ 0.019 &$-$0.004$\pm$ 0.008    &                      & 18 \\  
\hline                  
\end{tabular}
\end{center}
{\footnotesize $^{(1)}$Central oxygen abundance in the gas, as a result of the fit in the bulge region, calculated with the {\it O3N2} method.\\
$^{(2)}$See Figs.~\ref{grad-gas-estre-barras} \& ~\ref{grad-gas-estre-bulbos}\\
$\dagger$Same as Table~\ref{gradientes_INT_R_2}}
\end{table*}

\subsection{Comparison of the nebular gas abundances with the distribution of stellar metallicities}

While measurements of the ionised gas abundances give information about the current composition of the ISM, the 
stellar  metallicities are directly linked to the star formation history and, thus, a comparison between the metal abundances of both components can help us to determine the origin of the gas involved in the present star formation.

In previous works (P\'erez et al. 2007, 2009, 2011) we  derived the radial distribution of  stellar ages and metallicities along 
the bar and the bulges of the galaxies presented in this work.   Line strength indices were measured and used to derive 
age and metallicity gradients in the bulge and bar region by comparing with stellar population models. 

 Figure~\ref{com1} shows the comparison of the stellar and ionized gas abundances along the radius
for all  the galaxies. For the sake of clarity the radius is on a logarithmic scale and R=0 is shifted to log (R)=-0.5. For run2 galaxies the {\it R$_{23}$}(M91) calibration has been chosen, covering the error bars the values obtained with all methods with the exception of {\it O2Ne3}. The values obtained with all methods have been represented in the case of run1 galaxies. For 
the comparison, we take as reference the solar value 12+log(O/H)$_{\odot}$=8.69 (Allende-Prieto et al. 2001) and [Fe/H] for both components. It has been also shown the AGN character and we have marked with a vertical line the influence region of these active nuclei. In this way, we warn again (Sect. 4.2.1) the reader about comparing the gas abundances with the stellar metallicities in the regions dominated by an AGN.
\begin{figure*}
\centering
\includegraphics[width=\textwidth]{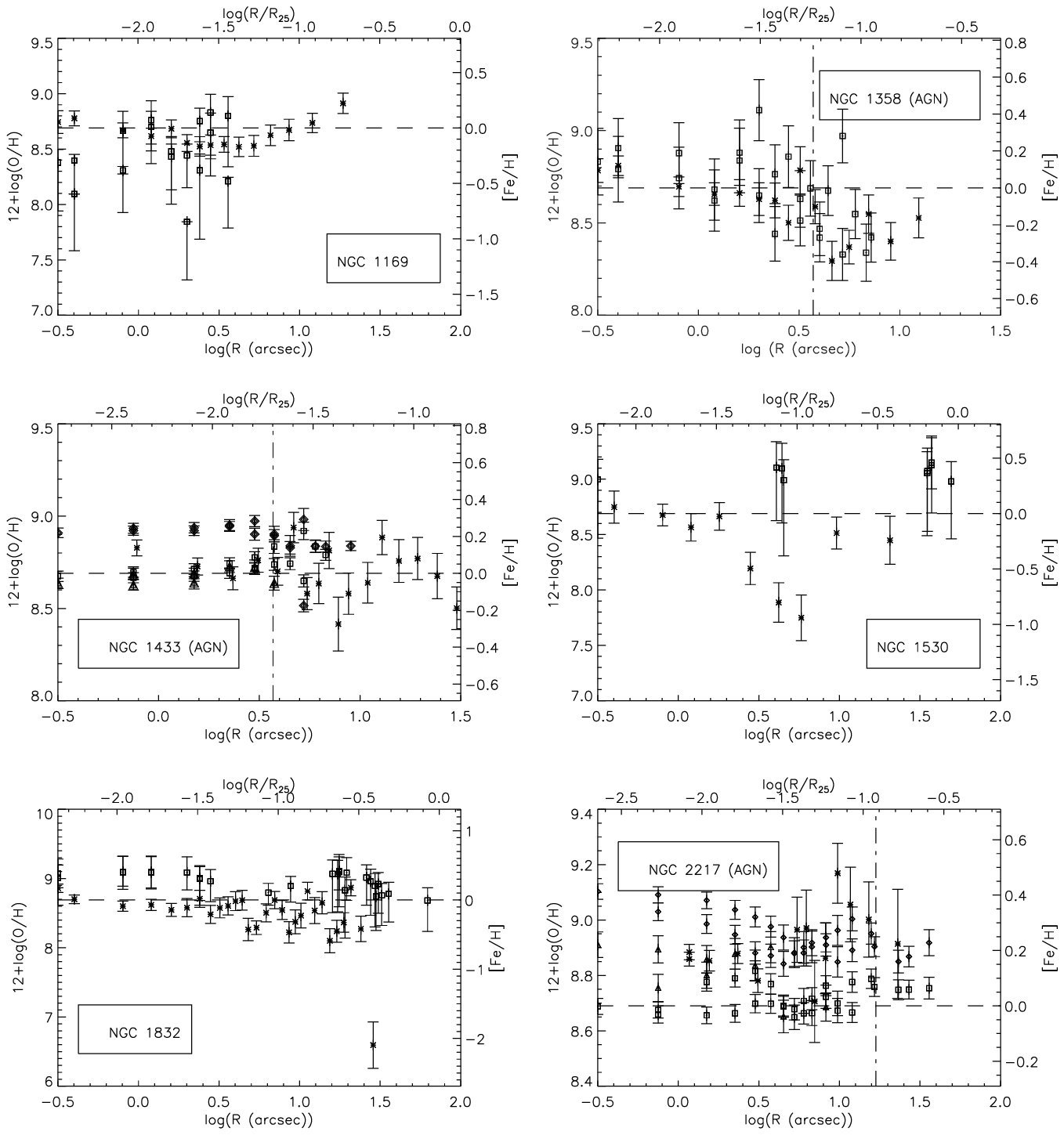}
\caption{Oxygen gas abundance and stellar metallicities relative to solar as a function of logarithm of galactocentric radius in arcsec. R$=0$ is shifted to $\log{R}=-0.5$. Asterisks indicate metallicities relative to solar obtained from the stellar population (see text for details). Open squares indicate oxygen gas abundances using {\it R$_{23}$} for run2 galaxies, and error bars cover values obtained with all methods but {\it O2Ne3}. Fon run1 galaxies, open squares indicate those obtained with {\it O3N2}, diamonds with {\it N2} and triangles with {\it SB1}. AGN galaxies are marked, with the vertical lines indicating the influence region of the active nuclei (either as in Fig.~\ref{actividad} or 3'', as indicated in references in V\'eron-Cetty \& V\'eron (2006)).}
\label{com1}%
\end{figure*}
\newpage
\begin{figure*}
\setcounter{figure}{10}
\centering
\includegraphics[width=\textwidth]{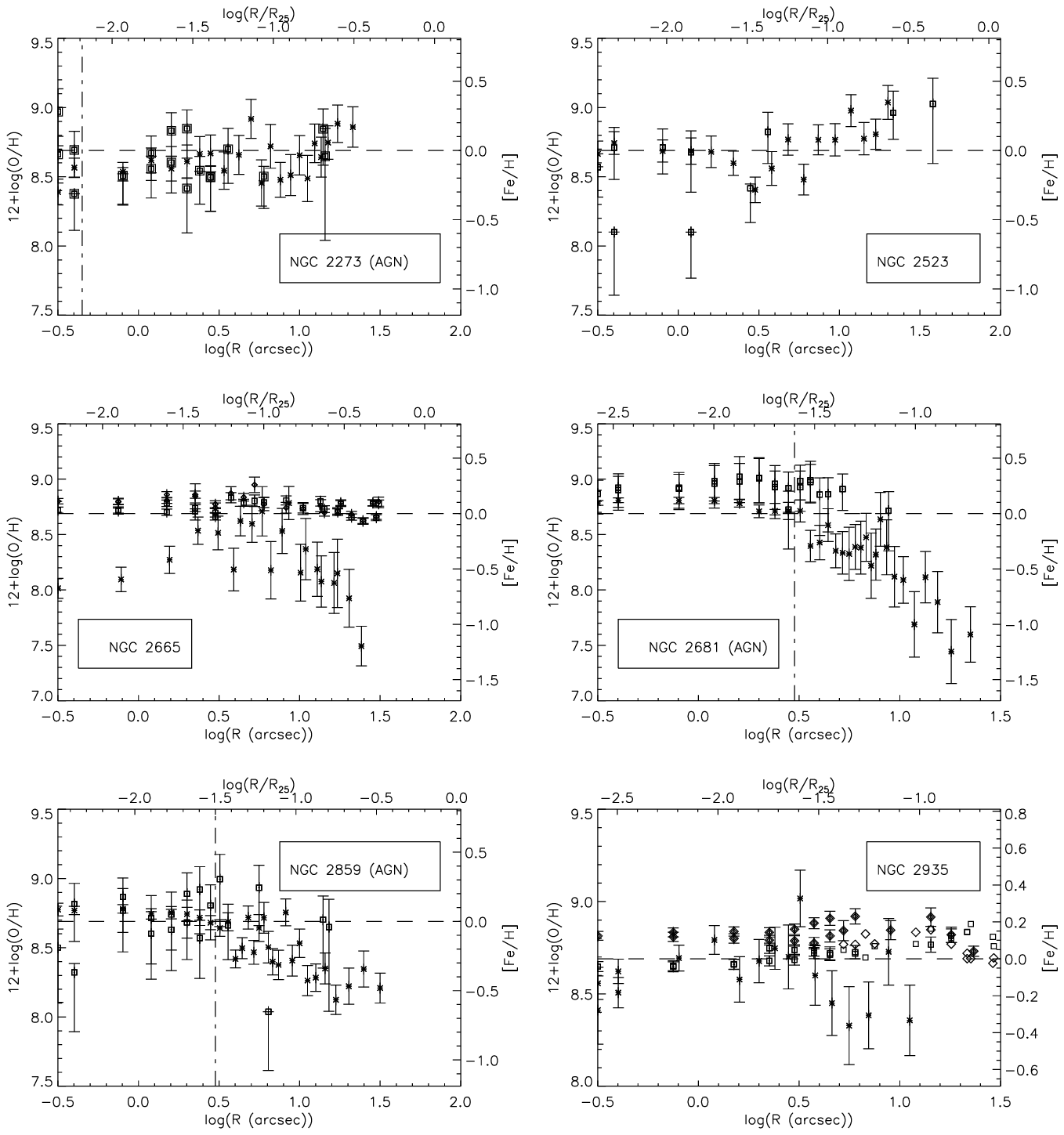}
\caption{Continued.}
\end{figure*}
\newpage
\begin{figure*}
\setcounter{figure}{10}
\centering
\includegraphics[width=\textwidth]{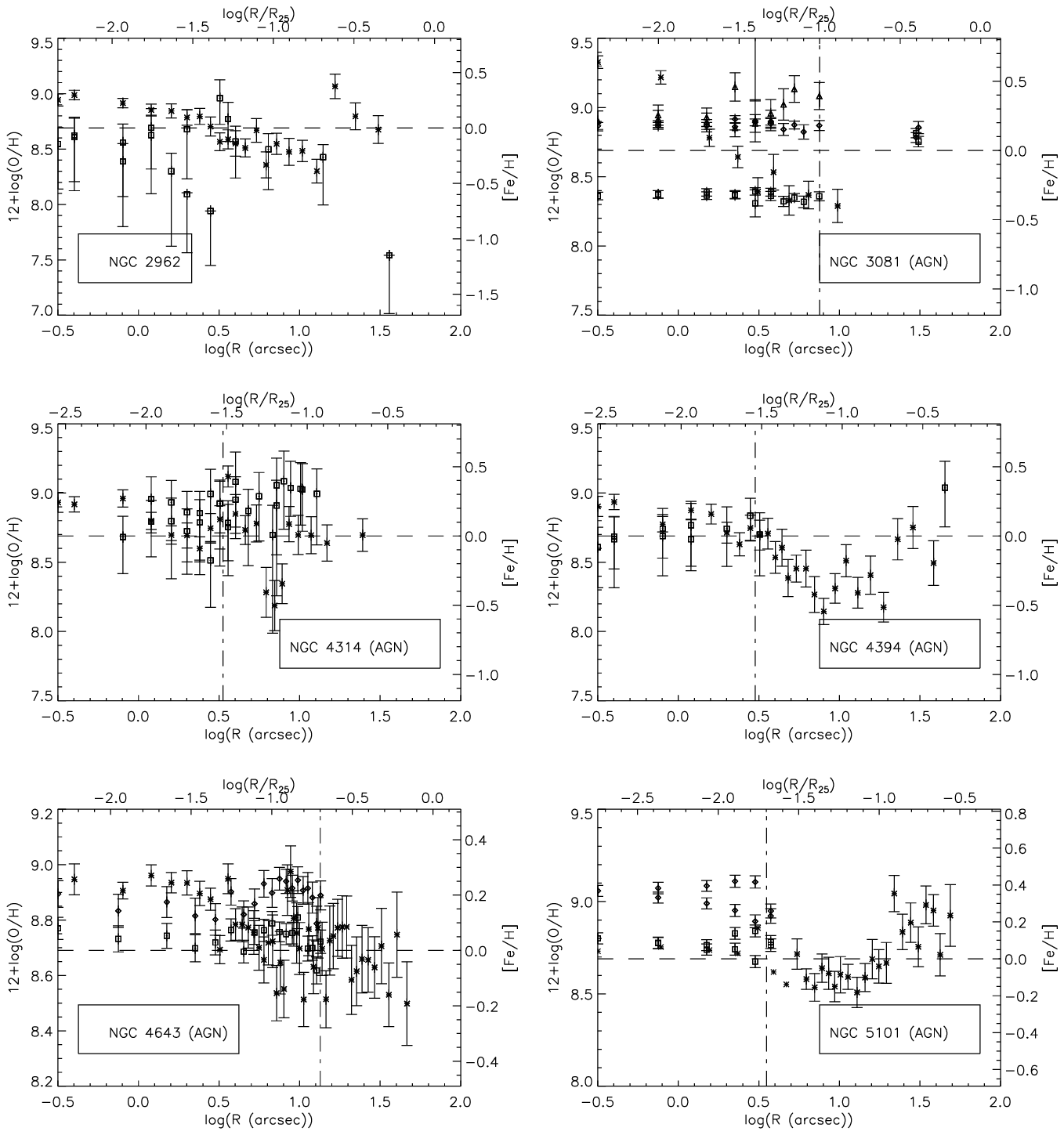}
\caption{Continued.}
\end{figure*}

For the following run2 galaxies the nebular metallicities lie clearly below the stellar metallicity values: NGC~1169, NGC~2962 (in agreement with  Marino et al. 2011) and NGC~4394. Only NGC~4394 is classified as LINER. This lower metallicity gas (w.r.t. the stellar metallicity) is located in the bulge region, similar values for the metallicity of both components  are found elsewhere. For run1 galaxies nebular metallicities obtained with {\it O3N2} are smaller than stellar metallicities in NGC~2217, NGC~3081 and NGC~4643. But with the values obtained with {\it N2} they are very similar (NGC~3081 and NGC~4643) or even higher (NGC~2217).  These three galaxies are classified as AGNs. If we consider {\it SB1}, nebular metallicity  is of the same order than stellar metallicity for NGC~3081 (Sy) and NGC~2217 (LINER).

For some of the other galaxies the distribution of the gas and the stellar abundances closely follow each other and for 3 galaxies 
the gas abundances lie clearly ($\ge$0.2 dex) above the stellar metallicities.

Not many works have addressed a direct comparison between the metallicities obtained from the stars and the ionized gas. Storchi-Bergmann et al. (1994)\
 concluded that both abundance measurements are well correlated for star-forming galaxies.
However, recent work (Marino et al. 2011) suggests, as derived from H{\small I}
and GALEX UV imaging,  the presence of current star formation activity in early
type galaxies with emission lines. This sort of galaxy rejuvenation could be due to
external gas re-fueling. This accretion
would suggest differences in the metallicities of the old stellar
component and the newly accreted gas, responsible for the present
star formation.  Furthermore, Annibali et al. (2010), found in a
comparison between nebular and stellar metallicities of E-S0 galaxies
with nebular emission that the gas metallicity tends to be lower that
the stellar metallicity, being the effect most severe for the largest
stellar metallicities.

\begin{figure}
   \centering
   \includegraphics[width=\columnwidth]{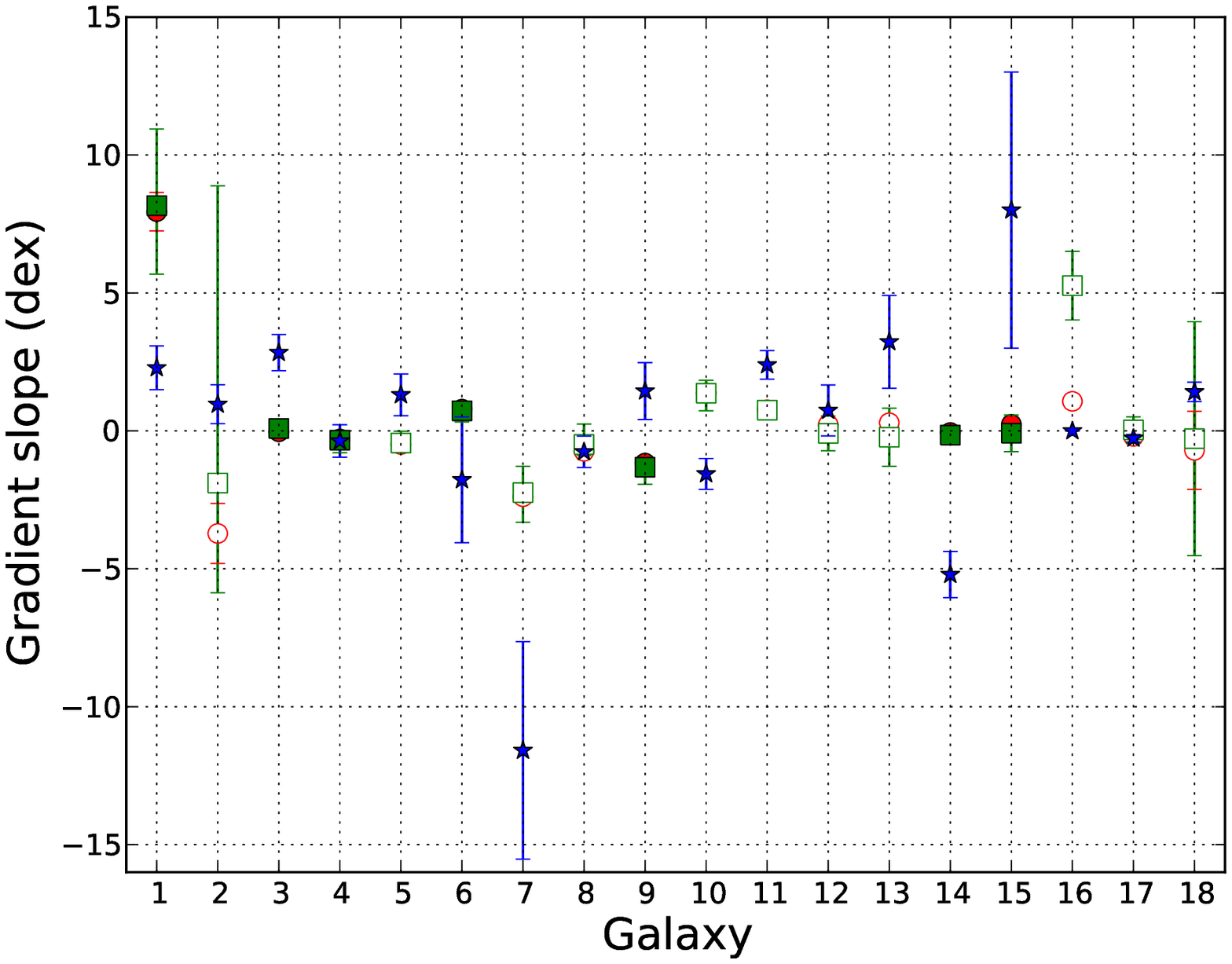}
   \caption{Comparation between metallicity gradients obtained for gaseous and stellar component in the bar region of our sample galaxies. They are normalized to R$_{25}$ (i.e. the gradients have been obtained from linear fits to the 12+log(O/H) {\it versus} R/R$_{25}$ representation for each galaxy). Circles are those estimated with the {\it R$_{23}$}(M91) or {\it N2} methods, squares are the mean value of gradients calculated with all methods (the error bar covers the range of these gradients),  and stars are the gradient [Z/H] for the stellar component (Paper II). Order in abcisas is that of Tables~\ref{gradientes_INT_bulbo} and \ref{gradientes_SSO_bulbo}. Filled symbols are for star-formation galaxies.}
              \label{grad-gas-estre-barras}%
    \end{figure}

\begin{figure}
   \centering
   \includegraphics[width=\columnwidth]{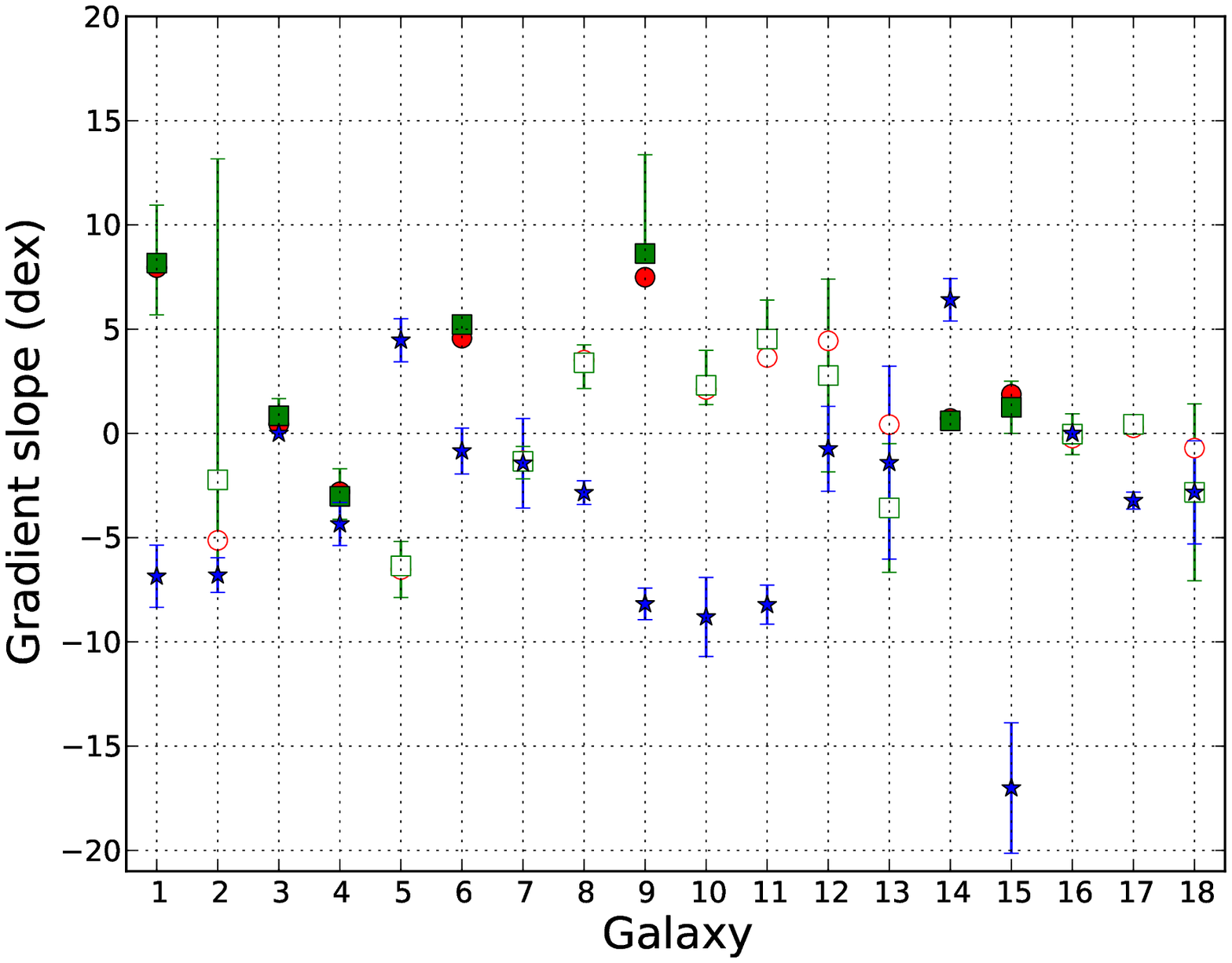}
   \caption{Comparation between metallicity gradients obtained for gaseous and stellar component in the bulge region of our sample galaxies. They are normalized to R$_{25}$ (i.e. the gradients have been obtained from linear fits to the 12+log(O/H) {\it versus} R/R$_{25}$ representation for each galaxy). Circles are those estimated with the {\it R$_{23}$}(M91) or {\it N2} methods, squares are the mean value of gradients calculated with all methods (the error bar covers the range of these gradients),  and stars are the gradient [Z/H] for the stellar component (Paper II). Order in abcisas is that of Tables~\ref{gradientes_INT_bulbo} and \ref{gradientes_SSO_bulbo}. Filled symbols are for star-formation galaxies.}
              \label{grad-gas-estre-bulbos}%
    \end{figure}

In Figs.~\ref{grad-gas-estre-barras} \& \ref{grad-gas-estre-bulbos} we compare the gradients obtained for the oxygen abundances along bars and the bulge regions with those estimated 
in Paper~II. They are normalized to $R_{25}$  ((12+log(O/H))/(R/R$_{25}$)).
In Paper~I we found that the stellar metallicity gradients along the bar show a large variety.  The distribution of the gas nebular abundances in the bar region shows a mild gradient (flatter than for the stellar component  for star-formation galaxies except for NGC~1169), and we do not see any clear relation between the stellar and nebular gas metallicity gradients. 
With respect to the bulge region, gradients for the stellar component in star-formation galaxies are flatter than the gaseous gradients for NGC~1169 and NGC~2523.
In the measurable cases, the oxygen abundance gradients are, in general, steeper for bulges than for bars, and they have the same sign. However, this is not the case for the stellar abundance gradient, where the sign between the bulge and the bar gradient changes in 10 out of 16 galaxies.

\section{Discussion}

Most of the emission is centrally 
concentrated in the bulge region. This has been noted before  and seems 
to be independent of the presence of a bar (Davis et al. 2011).
In 5 galaxies, emission lines are present along the whole extension of the bar, 
and 6 out of the 18 galaxies present some star formation at the bar ends as well 
as in the bulge region. Previous works have  claimed that the distribution of star 
formation in a bar is correlated with the bar age; younger bars ($<$~1Gyr) show star 
formation along the entire bar while older bar stages  show star formation at the bar-ends
 and nuclear region (Phillips 1993; Martin \& Friedli 1997; Martin \& Roy 1995; 
Friedli \& Benz 1995; Verley et al. 2007). However, some of the galaxies for which nebular gas 
is found along the bar seem to host bars that are old (see Paper I). Therefore, the star formation distribution in these 
bars is possibly more related to the reservoirs of gas that could replenish the bar than 
to the actual time of bar formation.
 
We do not find neither a relation between the difference in abundance values and the type of the galaxy 
nor a relation with the galactocentric distance in which the lines are measured (see Sect. 4.2.1 and 4.2.2).

We have considered the possible systematic effect of the impact of an AGN nuclei in the determination 
of abundances.  Its effect is difficult to quantify, but very important when comparing with the 
stellar metallicities in the inner regions of these galaxies and deriving the abundance gradients. 
The central regions will be a mixture of different components with different physical properties 
(Lindblad \& Fathi 2011) and will introduce difficulties in determining the main ionising mechanisms. 
Most of the central abundances  in the  works above mentioned, have not considered the influence 
of an active nuclei. Their abundance estimations have been carried out in \hii\ regions or/and in 
star formation galaxies, so the  models for calibration are based on stellar photoionization alone.  
Storchi-Bergmann et al. (1998)  proposed two calibrations for active galaxies: the first one depending 
on  [\nii]/H$\alpha$ and [\oiii]/H$\beta$ (SB1), and the second, a linear combination of 
log [\nii]/H$\alpha$ and [\oii]/[\oiii] (SB2). They concluded that, when possible, the best is to use 
the mean value of both.  Based on the extrapolation of metallicities towards the nuclear regions, 
they obtained that this calibration is good for Seyferts, but there is no clear conclusion for
 LINERs. We have used this method, SB1, only for three galaxies of run1. We cannot estimate SB2 for them 
because we cannot measure  [\oii]. For these galaxies, 2 LINERs and a Sy, we compare abundances calculated with {\it SB1} and {\it N2} and {\it O3N2}, that would be the extreme cases: i.e. to consider AGN or photoionization produced by hot stars. For LINERs, the results of {\it SB1} matches or is closer to that of {\it O3N2} in the central regions, while for the Seyfert galaxy they are similar to those estimated with {\it N2}. In any case, the difference between oxygen abundance 
using {\it N2} or {\it SB1} is less than 0.2 dex, a value considered as the intrinsic uncertainty of the models 
used to calibrate {\it R$_{23}$}. Although the presence of an AGN does not seem to alter the results, 
we would like to warn the reader about the validity of the results in the regions influenced by active nuclei.

We compare oxygen abundances with the stellar metallicities. We find central 
metallicities, 12+log(O/H), with a mean value within 1.2 arcsec, in the range of 8.4 to 9.1. These values agree with those found in the inner 
parts of early-type spirals.
Oey \& Kennicutt (1993) studied \hii\ regions in 15 Sa  to Sb galaxies, obtaining abundance values, 
with {\it R$_{23}$}, from 8.7 to 9.2. The larger values correspond to those regions located at 
lower galactocentric radii.  They find that  gas metallicities in these galaxies are systematically 
higher than those in Sc and later type galaxies. Zaritsky et al. (1994) discuss that this relation 
might reflect a dependence with galaxy mass and not with galaxy type after analysing  \hii\ regions 
in  a sample of 39 galaxies covering morphological types from Sab to Sd. They found metallicities 
in the range  8.2-8.6, similar  to the abundances of our work, although we cover morphological types 
from SB0 to SBbc.  Vila-Costas \& Edmunds (1992) also find a correlation between central gas abundances 
and galaxy mass, analysing  the gas metallicity distribution of 32 discs of different morphological 
types.  For the inner radii of 
Sab-Sb they find oxygen abundances in the range 8.7-9.5. A large statistical study using SDSS data by Tremonti et al. (2004) analyzes 53.000 star forming galaxies at z $\sim$ 0.1. In this study, the authors address the fundamental role that the galaxy mass plays in its chemical evolution. They find a strong correlation between mass and nebular abundance for galaxies with masses between  $10^{8.5}$  and $10^{10.5} M_\odot$, with a flattening of the relation for larger galaxy masses. Some recent studies investigating how mergers and bars can modify this correlation (Ellison et al. 2011) find a larger nebular abundance in barred galaxies than in unbarred ones of the same mass. 
Our galaxies lie on the flat part of the correlation, and therefore we do not expect any strong correlation between the rotational velocity and the nebular abundance.

\begin{figure}
   \centering
   \includegraphics[width=\columnwidth]{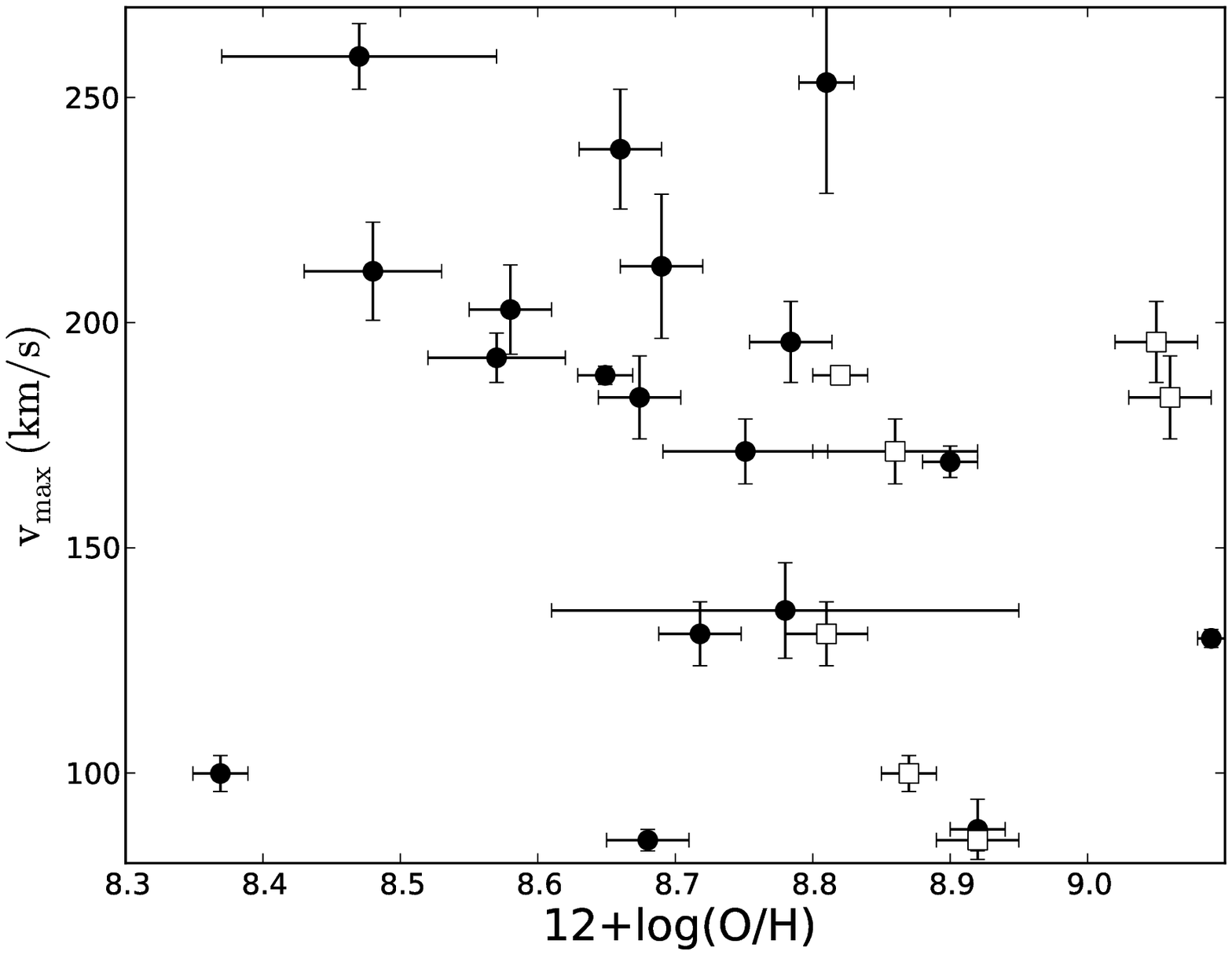}
   \caption{Rotational velocity corrected for inclination versus central oxygen abundances (1.2 arcsec). Filled circles represent those calculated using the {\it R$_{23}$} method for run2 and {\it O3N2} for run1. Squares are  values obtained with {\it N2} for run1.}
              \label{abun_vel}%
    \end{figure}

However, we see in Fig.~\ref{abun_vel} that the galaxies with the lowest abundances are those with the largest rotational velocities, although there is a large dispersion in the values, also finding galaxies 
with high rotational velocity and large ionized gas metallicities. However, there are no galaxies with low (12+log(O/H) $<$ 8.68) nebular abundances and low rotational velocities. The only exception is NGC~3081 (Seyfert), but with the abundance value obtained using {\it O3N2}. This result might be reflecting that it cannot be rejected some rejuvenation mechanism for some of the most massive galaxies in our sample.

\begin{figure}
   \centering
   \includegraphics[width=\columnwidth]{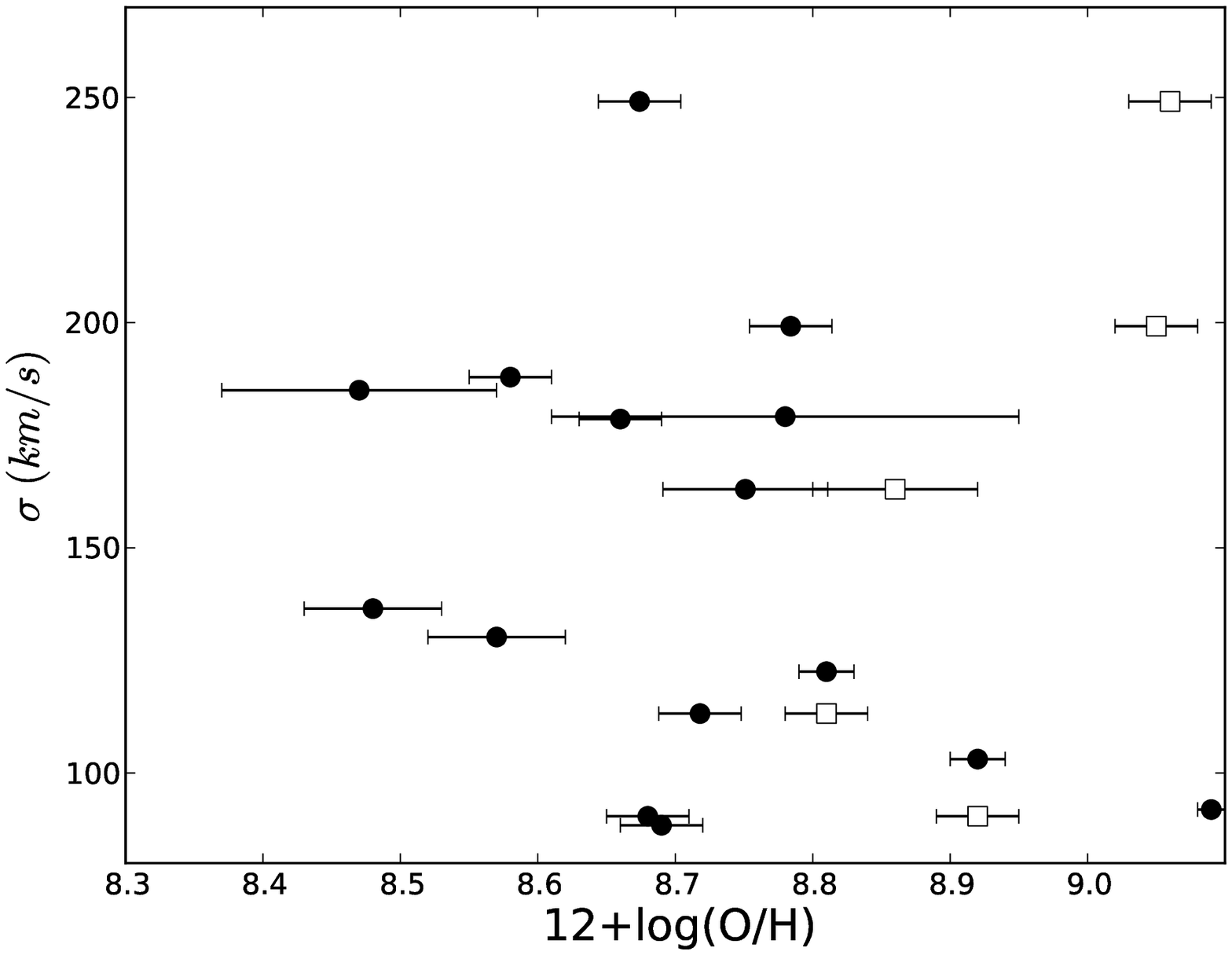}
   \caption{Stellar velocity dispersion versus central oxygen abundances (1.2 arcsec). Filled circles represent those calculated using the {\it R$_{23}$} method for run2 and {\it O3N2} for run1. Squares are  values obtained with {\it N2} for run1.}
              \label{abun_sigma}%
    \end{figure}

Following this result on the tendency of nebular abundances with disk rotational 
velocities, we search for nebular abundance trends with stellar velocity dispersion; however, we find 
no such trend.  As for the stellar abundances, we find in paper II, as expected from other studies, a correlation between stellar central abundances and central velocity dispersion (see Fig.~\ref{abun_sigma}).

Regarding the comparison of these central nebular abundances with central stellar metallicities, 
we find that for three of the galaxies the nebular abundances lie well bellow the stellar metallicities. 
The galaxies which show nebular abundances lower than the stellar 
metallicities are systematically among those galaxies with the largest rotational velocities of our sample. 

The bar is thought to be an efficient mechanism to quickly redistribute material within the disk. 
External accretion via the bar potential is a plausible mechanism 
to bring new gas to  the center and dilute the central gas metallicities. The strength of the bar 
determines the gas inflow rate (e.g. Athanassoula 1992, 2000); however, 
no clear trend between the bar strength and nebular composition is found for our galaxies. 
The efficiency of this process might be limited to the availability of gas, and although no trend 
with HI mass is found we do not posses a complete information about the total gas content of our galaxies.
Other processes capable of explaining the existence of low metallicity gas in the center of these 
massive galaxies include cold flows to the center (Kere{\v s} et al. 2005) and particular star formation 
histories (Yates et al. 2011). Yates et al. (2011) conclude, by examining semi-analytical models, 
that massive galaxies with low nebular abundances have undergone a gas-rich merger with a later shut-down 
of star formation. This low density metal-poor gas accretion is not sufficient to form stars efficiently 
and subsequently dilutes the gas-phase metallicities. Further information of the total gas content, its distribution and kinematics would help to elucidate the different external accretion mechanisms that might be at play in these galaxies.

As for the nebular abundance gradients found in this work, numerical models of barred galaxies (Friedli et al. 1994) show that the initial slope of the 
abundance is only slightly modified in the bar region, while outside the bar corotation radius, the 
original slope gets flatten. We have  estimated the gas oxygen abundances along the bars to study the 
efficiency of mixing by the bar inside corotation. Considering only star-forming galaxies, we find positive gradients, with a 3-$\sigma$ significance, for the oxygen abundance in the bar region in three galaxies and negative gradients in two of the 
galaxies. The estimated gradient values range from -3.7  to 1.4, in units of 
dex/(R/R$_{25})$ with the mentioned exception of NGC~1169. In 8 of the observed 
galaxies the gradient is lower than 0.5 dex/(R/R$_{25}$).  In the bulge region, where 
most of the emission is found, we find steeper gradients.

It is, however, not a simple task to interpret the origin of these gradients because not only stellar 
evolution, star formation and bar dynamics, already complicated processes, affect the gradients but 
also refueling of newly accreted gas could change the gradient. Naively, central accretion of gas would 
dilute the inner regions generating a positive gradient, while the effect of radially accreted gas could 
be not so evident without proper modelling. However, it would be possibly to reduce a pre-existing  gradient because of the rapid mixing and reactivation of star formation in the bar region (Friedli et al. 1994, 1995). 

Most of the gas metallicity gradients that can be found in the literature include the whole disk; 
van Zee et al. (1998) found gradients in the range [$-0.30$, $-1.52$] dex/R$_{25}$ for spiral galaxies, and Rupke et al. (2010) a mean value of -0.57 $\pm$ 0.05 considering 11 isolated spirals. 
Oey \& Kennicutt (1993) concluded that barred galaxies have a shallower gradient than unbarred ones. But a large range of abundance gradients have been found, some of them being much more steeper than for unbarred spirals (Edmunds \& Roy 1993, Considere et al. 2000).
Vila-Costas \& Edmunds, found gradients raging from $-3.33$ to $-0.899$ for non-barred galaxies, and from
$-0.156$ to $-0.398$ for barred galaxies (all in dex/R$_{25}$). Zaritsky et al. (1994) obtained a mean 
value of $-0.23$ dex/R$_{25}$ considering five barred galaxies. In particular, for the Milky Way, 
Balser et al. (2011) find a radial gradient of $-0.0446$ $\pm$ 0.0049 dex/kpc. Advardsson (2002) 
found a value of $-0.07$ $\pm$ 0.01 dex/kpc for galactocentric distances between 6 and 18 kpc.

The discrepancy between the values found in the literature for the disk and our values, concerning 
only the bar region, could be due to the predicted change in the slope around the bar radius 
(Friedli et al. 1994). Furthermore, we find a slope change between the bulge and the bar region, 
also predicted by Friedli et al. (1994).  
These changes in the abundance slope with radius, beeing steeper in the inner parts of the galaxies, 
have been analised by other authors. Zaritsky et al. (1994) found different gradient in the bar of 
NGC~3319 than outside of it. This question has been studied in later works, trying to detect a radial 
variation of the gradient. For instance, for the barred galaxy NGC~3359 Martin \& Roy (1995) found a 
radial variation of the gradient being, in turn, studied by Zahid \& Bresolin (2011), who estimated a 
break at a characteristic radius. Balser et al. (2011) concluded that there was no radial 
discontinuities for the Milky Way, but that the gradient varies in the azimuthal coordinate between 
$-0.03$ and $-0.07$ dex/kpc.  

Previous works  have assessed the effect of the AGN presence in the nebular abundance gradients. Annibali et al. (2010) showed that the nebular abundances, derived with the  
{\it R$_{23}$} method,  increase with radius, while they decrease when using methods that take into 
account harder ionising sources (i.e AGN), such as  the SB method, a combination of SB1 and SB2.
We have checked whether our gradient calculation could be biased in the same way. We can estimate SB1 for three run1 galaxies, NGC~1433 and NGC~2217 and NGC~3081. For them, we have not [\oii], so we cannot calculate {\it R$_{23}$}, but {\it N2} and {\it O3N2}. We find that {\it SB1} gradients are nearly flat for NGC~1433 and NGC~3081, and have the same sign than that from {\it N2} for NGC~2217.

There have been studies correlating the observed nebular metallicity gradients with physical properties 
of the galaxies. Zaritsky et al. (1994) suggested a correlation between the gradient and the 
bar-type as previously noticed by Pagel \& Edmunds (1981), Martin (1992) and Edmunds \& Roy (1993). We have not found any correlation between the estimated gradients and the physical properties of galaxies: e.g.; bar-type, HI mass or central stellar velocity dispersion. However, we cover a short range of masses and morphological types. Further studies covering a larger range of galaxy parameters and spatial range will be necessary to draw strong conclusions about the correlation between nebular abundance gradients and the physical properties of galaxies. 

\section{Summary and conclusions}

We have carried out a detailed analysis of the nebular abundances along
the bar and in the bulge of a sample of 20 early--type galaxies to compare them with 
the stellar abundance distributions obtained in previous works using 
line-strength indices. We focus on several relations, in particular, the distribution of abundances and star formation and the properties of the galaxies; the interpretation of the ionizing mechanisms and the relative influence of the AGN; the relation between abundances and rotational velocities and velocity dispersion in the bulge. 
We have found that most of the emission in our sample of galaxies are concentrated in the bulge region. 
We have used several methods to estimate nebular abundances. Considering different methods for star-forming galaxies, does not change our results about gaseous metallicity gradients nor change the comparison with the stellar metalicities. In three galaxies the estimated nebular abundances lie clearly below the stellar metallicities.  Furthermore, we see that the galaxies with the lowest abundances are those with the largest rotational velocities, and although there is a large dispersion in the values, this effect might be the result of a central rejuvenation mechanism in the most massive and late-type galaxies. The comparison between gaseous and stellar component shows that in most galaxies the slope for the nebular abundance distribution in the bulge region is shallower than that of the stellar metallicity.
We can conclude that the combination of observations of both gas and stellar metallicities is crucial to determine the star formation history in galaxies. The study of both phases is very useful to learn about the origin of the gas involved in the star formation; for instance, we clearly see galaxies for which external accreted gas is the fuel source for the present star formation.  Due to the small parameter range covered by our sample we can not conclude anything about the role of bars in the mixing, but we observe a variety of gas and stellar metallicities distributions that indicate very different mixing and inflow histories from galaxy to galaxy. Therefore, it would be very interesting to complete this study with a wider sample, including later-type and unbarred galaxies.

\begin{acknowledgements}
We are grateful to the referee for the useful comments and suggestions that have improved the manuscript.
We acknowledge the usage of the HyperLeda database (http://leda.univ-lyon1.fr). This research has been supported by
the Spanish Ministry of Science and Innovation (MICINN) under grants
AYA2011-24728, AYA2010-21322-C03-02, AYA2010-21322-C03-03, AYA2007-67625-C02-02 and Consolider-Ingenio CSD2010-00064, and 
by the Junta de Andaluc\'ia (FQM-108). PSB acknowledge support
from the Ramon y Cajal Program ﬁnanced by the Spanish Ministry of Science
and Innovation. PSB also acknowledges an ERC grant within the 6th European
Community Framework Programme. 
\end{acknowledgements}

\end{document}